\documentclass[twocolumn, amsmath, amssymb, aps, pre, superscriptaddress]{revtex4-1}
\usepackage{xcolor, mathtools, graphicx, dcolumn, bm, hyperref}

\def\be{\begin{equation}}
\def\ee{\end{equation}}
\def\ba{\begin{eqnarray}}
\def\ea{\end{eqnarray}}

\begin{document}


\title{Proof of the zeroth law of turbulence in one-dimensional compressible magnetohydrodynamics and shock heating}

\author{Vincent David}
\affiliation{Laboratoire de Physique des Plasmas, \'Ecole polytechnique, F-91128 Palaiseau Cedex, France}%
\affiliation{Universit\'e Paris-Saclay, IPP, CNRS, Observatoire Paris-Meudon, France}%

\author{S\'ebastien Galtier}
\affiliation{Laboratoire de Physique des Plasmas, \'Ecole polytechnique, F-91128 Palaiseau Cedex, France}%
\affiliation{Universit\'e Paris-Saclay, IPP, CNRS, Observatoire Paris-Meudon, France}%
\affiliation{Institut universitaire de France}%

\date{\today}

\begin{abstract}
The zeroth law is one of the oldest conjecture in turbulence that is still unproven. Here, we consider weak solutions of one-dimensional compressible magnetohydrodynamics and demonstrate that the lack of smoothness of the fields introduces a new dissipative term, named inertial dissipation, into the expression of energy conservation that is neither viscous nor resistive in nature. We propose exact solutions assuming that the kinematic viscosity and the magnetic diffusivity are equal, and we demonstrate that the associated inertial dissipation is positive and equal on average to the mean viscous dissipation rate in the limit of small viscosity, proving the conjecture of the zeroth law of turbulence and the existence of an anomalous dissipation. 
As an illustration, we evaluate the shock heating produced by discontinuities detected by Voyager in the solar wind around 5\,AU. We deduce a heating rate of $\sim 10^{-18}$J\,m$^{-3}$\,s$^{-1}$, which is significantly higher than the value obtained from the turbulent fluctuations. This suggests that collisionless shocks can be a dominant source of heating in the outer solar wind.
\end{abstract}


\maketitle

\section{\label{sec:intro}Introduction}
One of the oldest problems in turbulence is the so-called zeroth law which states that if in a turbulent flow experiment all control parameters are fixed except viscosity, which is lowered as much as possible, the mean energy dissipation per unit mass tends towards a non-vanishing limit independent of viscosity (Onsager's conjecture) \cite{Onsager49,Frisch95,Eyink94,SaintMichel2014}. For magnetohydrodynamic (MHD) fluids, this law is mathematically translated as follows
\be \label{eq:0thlaw}
    \lim_{\nu, \eta \to 0^{+}} \frac{\mathrm{d}E}{\mathrm{d}t} = - \varepsilon < 0 ,
\ee
where $E$ is the mean total energy, $\nu$ the viscosity, $\eta$ the magnetic diffusivity and $\varepsilon$ the mean rate of total energy dissipation (per unit mass) \cite{G16}. The conjecture (\ref{eq:0thlaw}) has been relatively well verified with 3D direct numerical simulations in isotropic \cite{Mininni09} and non-isotropic \cite{Bandyopadhyay18} MHD cases. The zeroth law is fundamental because it is a basic assumption made to derive exact laws in hydrodynamics \cite{K41,Monin59,Antonia97,Lindborg1999,G11,F20}, MHD \cite{PP98a,BG13,Wan2009} or Hall MHD \cite{G08,Andres18,F19} turbulence. 

Space plasmas (solar wind, Earth's magnetosheath) are often in a regime of fully developed turbulence. For the solar wind, a question remains open: knowing that it can be considered as an expanding isolated system, the evolution of its temperature $T$ with the heliocentric distance $r$ should be close to an adiabatic law with $T(r) \sim r^{-4/3}$. However, {\it in situ} measurements reveal that the solar wind is warmer than expected with a slow decrease in temperature (close to $T(r) \sim r^{-1/2}$ for $r <10$\,AU), reflecting the presence of a local heating \cite{Gazis1982,Richardson1995,Matthaeus99}. On the other hand, the extremely high (magnetic) Reynolds numbers ($\ge 10^{10}$) make turbulence a potential candidate to explain the presence of an efficient heating fed by turbulence. This question has motivated many studies to estimate space plasma heating by {\it in situ} measurements of $\varepsilon$ 
\cite{sorriso2007,Vasquez07,macbride08,Marino08,Carbone09,Stawarz09,coburn15,Banerjee2016,Hadid2018,Andres2019}. 

The measurement of $\varepsilon$ in space plasmas is performed using exact laws that implicitly assume statistical homogeneity (which means stationary time interval with the Taylor hypothesis).
However, the velocity and magnetic field fluctuations are not always statistically homogeneous:
in \cite{Hadid2017} (see Fig. 17) it is shown that the presence of discontinuities breaks this property, which eventually distorts the estimate of $\varepsilon$ and then makes the use of exact laws less reliable. This effect can be removed by selecting intervals free of discontinuities. This 'filtering' (or selection of intervals) is often done for this type of study, but it is not always explicitly mentioned. The use of a local (\textit{i.e.} non-statistical) approach allows us to estimate the energy dissipation rate by freeing ourselves from the assumption of statistical homogeneity \cite{Duchon}. 
The local approach assumes that the fields are distributions (generalized functions) and may, therefore, present discontinuities. Solving fluid equations under this assumption is mathematically equivalent to searching for so-called weak solutions \cite{Leray34}. Coming back to the solar wind, the presence of discontinuities in the data is well-known in particular at 2--10\,AU \cite{Burlaga1984,Roberts1987}, which reinforces the interest to develop a local approach and to understand the ins and outs, especially since in hydrodynamics  \cite{Duchon,Dubrulle19} and MHD \cite{G18} it has been shown that the local approach introduces a new type of dissipation called inertial due to the lack of smoothness of the fields. 

Burgers' equation is often used as a one-dimensional (1D) model of turbulence because, in appearance, it has much in common with the three-dimensional (3D) Navier-Stokes equations \citep{Bec2007}. In particular, it can be used to analytically probe the essence of turbulence \cite{Frisch95}. An equivalent for MHD fluids has been proposed by \citet{Yanase97} (see also the \textit{ad-hoc} model proposed by \citet{Thomas1968}) which conserves both the total energy and a quantity similar to the cross-helicity. 1D MHD models are often used to investigate astrophysical problems like the turbulent dynamo, i.e. the growth of an initially weak magnetic field, within planets and stars \cite{Thomas1968}, the solar coronal heating in magnetic loops \cite{Galtier1998} or in coronal holes with a heating produced by shocks \cite{Suzuki2005}. 

In this article, we derive analytical properties (inertial dissipation, exact solutions and mean dissipation rates) of the Yanase equations from which we prove the zeroth law of turbulence (sections 2, 3, 4 and 5). In section 6, as an illustration we study the outer solar wind around 5\,AU where sawtooth and battlement structures are observed. The expression of the inertial dissipation provides a simple formula that is used to estimate for the first time the heating produced by such collisionless shocks. A conclusion is given in the last section.

\section{\label{sec:model} Yanase's equations}
We start from the 3D compressible MHD equations (with $\nabla \cdot \boldsymbol{B} = 0$)
\ba
    \partial_t \rho + \nabla \cdot \left( \rho \boldsymbol{u} \right) &=& 0, \label{eq0} \\
    \rho \left(\partial_t \boldsymbol{u} + \boldsymbol{u} \cdot \nabla \boldsymbol{u}  \right) &=& -\nabla P + \frac{1}{\mu_0} \left( \nabla \times \boldsymbol{B} \right) \times \boldsymbol{B} + \mathbf{d}_{\tilde\nu}, \,\, \label{eq1} \\
    \partial_t \boldsymbol{B} &=& \nabla \times \left( \boldsymbol{u} \times \boldsymbol{B} \right) + \mathbf{d}_\eta, \label{eq2} 
\ea
\noindent
where $\rho$ is the mass density, $\boldsymbol{u}$ the velocity, $\boldsymbol{B}$ the magnetic field, $P$ the pressure, $\mu_0$ the magnetic permeability of the vacuum, $\mathbf{d}_{\tilde \nu} = \tilde \nu \nabla^2 \boldsymbol{u} + (\tilde \nu /3) \nabla \left( \nabla \cdot \boldsymbol{u} \right)$ and $\mathbf{d}_\eta = \eta \nabla^2 \boldsymbol{B}$ are the dissipative terms, $\tilde\nu$ the dynamic viscosity and $\eta$ the magnetic diffusivity. \citet{Yanase97} reduces this system by making the following assumptions: $\boldsymbol{u}$ and $\boldsymbol{B}$ depend only on  the 1D space variable $x$ and time $t$ such that $\boldsymbol{u}(x,t) = u(x,t) \boldsymbol{e}_x$ and $\boldsymbol{B}(x,t) = B_y(x,t) \boldsymbol{e}_y + B_z(x,t) \boldsymbol{e}_z$. 
The pressure term is neglected compared to the magnetic pressure (small $\beta$ limit); the density $\rho$ is put constant and equal to $\rho_0$ in the equations for the velocity and magnetic fields. All this leads to the Yanase equations
\ba
\label{eq:dudtYanase}
\partial_t u + u\partial_x u + b_y \partial_x b_y  + b_z \partial_x b_z  &=& \nu \partial_{xx} u,\\
\label{eq:dbydtYanase}
\partial_t b_{y} + u\partial_x b_y + b_y \partial_x u  &=& \eta \partial_{xx} b_{y},\\
\label{eq:dbzdtYanase}
\partial_t b_{z} + u\partial_x b_z + b_z \partial_x u  &=& \eta \partial_{xx} b_{z},
\ea 
\noindent
where by definition $b_j \equiv B_j / \sqrt{\rho_0 \mu_0}$ with $j=y,z$ and $\nu \equiv 4 \Tilde{\nu} / \left( 3 \rho_0 \right)$. It is straightforward to show that the total energy, $E = (u^2+b_y^2 + b_z^2)/2$, and the modified cross-helicity, $u b$ with $b$ the magnetic field modulus, are conserved when $\nu=\eta=0$. Although this 1D system mimics the compressible MHD equations with two similar invariants \citep{G16}, it is generally not fully consistent with it because of the assumption of constant density (in space and time), which leads necessarily to a constant velocity $u$. There is, however, one exception to this (considered in this article): when the velocity derivative is constant in space, the mass density will remain constant in space (but not in time) if it is initially so. 

\section{\label{sec:energy}Inertial dissipation}
In the classical picture of turbulence the fields $u$, $b_y$ and $b_z$ are assumed to remain smooth at all scales. If it is not the case, one needs to regularize Eqs. (\ref{eq:dudtYanase})--(\ref{eq:dbzdtYanase}). To do this, we introduce $\varphi$ an infinitely differentiable function with compact support on $\mathbb{R}$, even, non-negative with integral 1 \cite{Duchon}. We also define $\varphi^\ell(\xi) \equiv \varphi(\xi / \ell) / \ell$, where $\xi \in \mathbb{R}$ and $\ell \in \mathbb{R}_+^*$. Denoting $u^{\ell} = \varphi^\ell * u$, $b_y^{\ell} = \varphi^\ell * b_y$ and $b_z^{\ell} = \varphi^\ell * b_z$, one obtains (we consider a periodic domain)
\ba \label{eq:duldt}
    \partial_t u^{\ell} + \partial_x \left[\frac{(u^2)^{\ell} + (b_y^2)^{\ell} +(b_z^2)^{\ell}}{2} - \nu \partial_x u^{\ell} \right] &=& 0,  \\
    \label{eq:dbyldt}
    \partial_t b_y^{\ell} + \partial_x \left[ \left( ub_y \right)^{\ell} - \eta \partial_x b_y^{\ell} \right] &=& 0,  \\
    \label{eq:dbzldt}
    \partial_t b_z^{\ell} + \partial_x \left[ \left( ub_z \right)^{\ell} - \eta \partial_x b_z^{\ell} \right] &=& 0,
\ea
from which we can deduce
\be\label{eq:energy}
\begin{split}
    &\partial_t \left( \frac{u u^{\ell} + b_y b_y^{\ell} + b_z b_z^{\ell}}{2} \right) 
    + 
    \frac{u^{\ell}}{4} \partial_x \left( u^2 +b_y^2 + b_z^2 \right) \\
    &+
    \frac{u}{4} \partial_x \left[ \left(u^2\right)^{\ell} + \left(b_y^2\right)^\ell + \left(b_z^2\right)^\ell \right] \\
    &+
    \frac{b_y^{\ell}}{2} \partial_x \left( ub_y \right) + \frac{b_y}{2} \partial_x (u b_y)^{\ell}
    +
    \frac{b_z^{\ell}}{2} \partial_x \left( ub_z \right) + \frac{b_z}{2} \partial_x (u b_z)^{\ell} \\
    &- \partial_{xx} \left( \nu u u^{\ell} + \eta b_y b_y^{\ell}  + \eta b_z b_z^{\ell} \right) = - \mathcal{D}_{\nu, \eta}^\ell, 
\end{split}
\ee
where by definition
\be 
\mathcal{D}_{\nu, \eta}^\ell \equiv \nu \left(\partial_x  u \right) \left(\partial_{x} u^{\ell}\right) + \eta \sum_j \left(\partial_x  b_j\right) \left(\partial_{x} b_j^{\ell}\right) 
\ee
is the viscous/resistive dissipative term. 
The third-order structure functions are introduced in the following manner
\be \label{eq:inertial_dissipation}
    \mathcal{D}_I^\ell \equiv \frac{1}{12}\int \frac{\mathrm{d} \varphi^\ell}{\mathrm{d}\xi} \left[ \left(\delta u\right)^3 +3 \left( \left(\delta b_y\right)^2 + \left(\delta b_z\right)^2 \right) \delta u \right]  \mathrm{d}\xi,
\ee
\noindent
where $\delta u \equiv u(x+\xi) - u(x)$, $\delta b_y \equiv b_y(x+\xi) - b_y(x)$ and $\delta b_z \equiv b_z(x+\xi) - b_z(x)$. After an integration by parts and development, one finds
\begin{equation*}
    \begin{split}
\mathcal{D}_I^\ell = - \frac{1}{12} \left[
\partial_x (u^3)^{\ell} - 3u \partial_x \left(u^2 + b_y^2 + b_z^2 \right)^\ell \right. \\
\left. + 3 \left( u^2 + b_y^2 + b_z^2 \right) \partial_x u^{\ell} + 6 u \left( b_y \partial_x b_y^\ell + b_z \partial_x b_z^\ell \right) \right. \\
\left. + 3 \partial_x \left[ u \left( b_y^2 + b_z^2 \right) \right]^\ell - 6 b_y \partial_x \left( u b_y \right)^\ell - 6 b_z \partial_x \left( u b_z \right)^\ell \right] .
\end{split}
\end{equation*}
\noindent
Introducing the previous expression into the local expression of energy conservation (\ref{eq:energy}) leads to the point-splitting energy conservation equation
\ba
\begin{split}
&\partial_t \left( \frac{u u^{\ell} + b_y b_y^{\ell} + b_z b_z^{\ell}}{2} \right) + \partial_x \left[ \frac{(u^3)^{\ell}}{12} + \frac{u^{\ell} \left(b_y^2 + b_z^2\right)^{\ell} }{4} \right. \\
&\left. + \frac{u^{\ell} \left(u^2 + b_y^2 + b_z^2 \right)}{4} 
+ \frac{u \left( b_y b_y^{\ell} +  b_z b_z^{\ell} \right)}{2} \right] \\
&- \nu \partial_{xx} \left(u u^{\ell} \right) - \eta \partial_{xx} \left(b_y b_y^{\ell} + b_z b_z^{\ell} \right) = - \mathcal{D}_I^\ell - \mathcal{D}_{\nu, \eta}^\ell .
\end{split}
\ea
The limit $\ell \to 0$ leads to the first main result
\ba
\label{eq:dedt}
\partial_t E + \partial_x \Pi = - \mathcal{D}_I - \mathcal{D}_{\nu, \eta},    
\ea
with
\ba
\Pi &=& \frac{u^3}{3} + ub^2 - \partial_x \left(\nu u^{2} + \eta  b^2 \right), \\
\mathcal{D}_I &=& \lim_{\ell \to 0} \mathcal{D}_I^\ell, \\
\mathcal{D}_{\nu, \eta} &=& \nu \left( \partial_x u\right)^2 + \eta \sum_j \left(\partial_x  b_j\right)^2,
\ea
where $\Pi$ is the energy flux. Expression (\ref{eq:dedt}) is particularly relevant in the inviscid/ideal limit, \textit{i.e.} when $\nu = \eta = 0$. In this case, there is still a channel to dissipate energy through the inertial dissipation $\mathcal{D}_I$. This dissipation happens because of the lack of smoothness of the fields. On the contrary if the fields are regular enough, by using a Taylor expansion it is straightforward to show that $ \mathcal{D}_I^\ell \to 0$ when $\ell \to 0$. Finally, note that we recover the well-known result on Burgers' equation when b=0 \cite{EyinkNotes,Dubrulle19}.

\section{\label{sec:solution} Analytical solutions}
\begin{figure}
\includegraphics[width=\columnwidth]{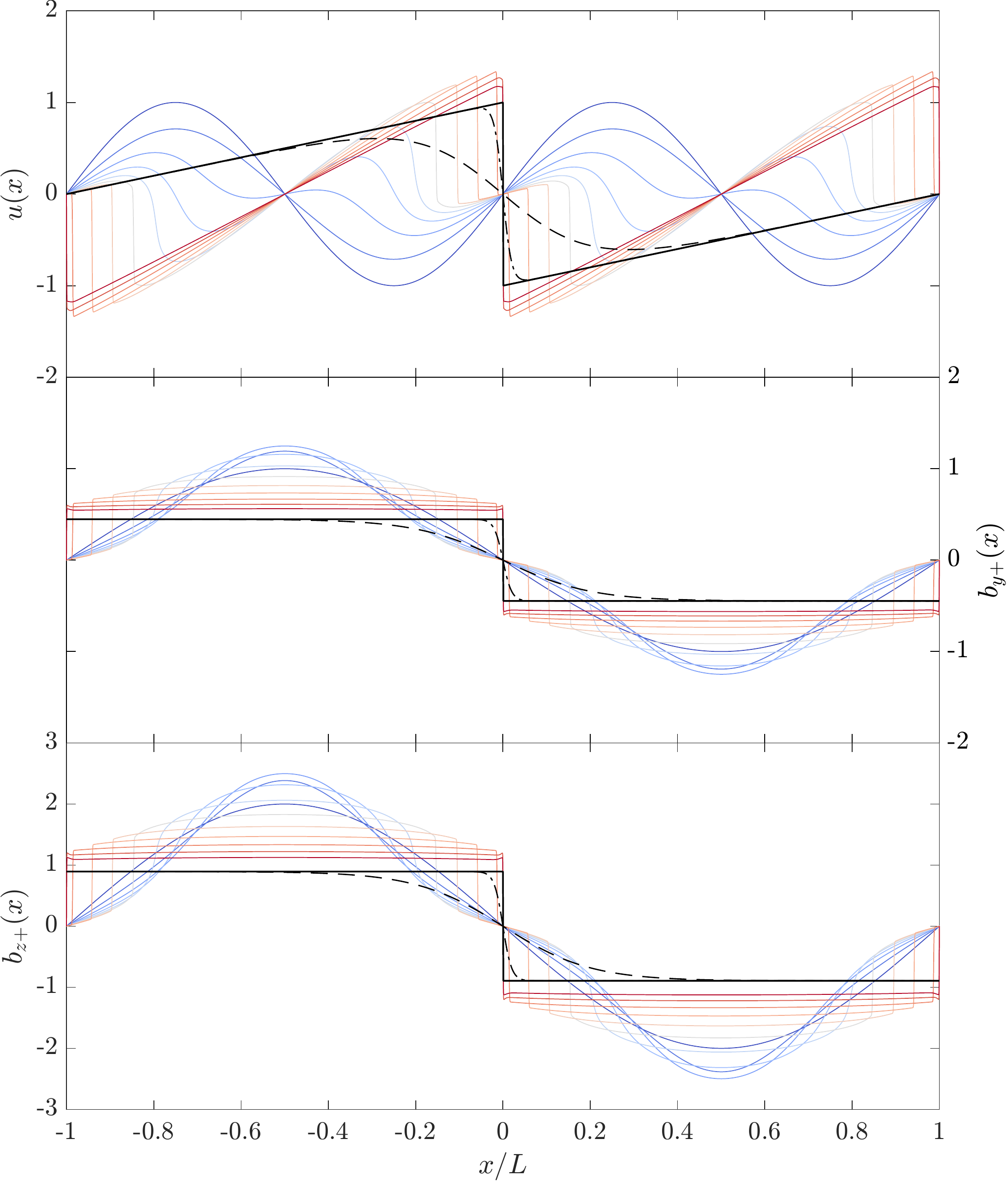}
\caption{Analytical solutions $u$ (top), $b_{y+}$ (middle) and  $b_{z+}$ (bottom) at $t=1$, for $\nu=0$, $0.01$ and $0.1$ (solid, dash-dot and dash black lines, respectively), in a domain of size $2L=2$. We take $c=2$ and thus the amplitude of the shocks are $\Delta=2$, $\Delta_y = 2/\sqrt{5}$ and $\Delta_z = 4/\sqrt{5}$. Superimposed: direct numerical solution of Yanase's equations with the time evolution (from blue to red) of the velocity (top) and the magnetic field components (middle and bottom).}
\label{fig:Yanase_Fields}
\end{figure}

Let us define the $u(x,t)$ and the magnetic field components ${b_y}_\pm(x,t)$ and ${b_z}_\pm(x,t)$ in the interval $x \in [-L,+L]$, with $L \in \mathbb{R_+}$ and $t\in \mathbb{R}_+^*$. If the dissipative coefficients $\nu = \eta \in \mathbb{R_{+}^*}$, then the Yanase equations admit the following (not necessarily unique) analytical solutions
\ba
\label{eq:solU}
u(x,t) &=& \frac{x}{t}-\frac{L}{t}\tanh{\left(\frac{xL}{\nu t}\right)}, \\
\label{eq:solBy}
{b_y}_\pm(x,t) &=& \pm b_{y,0} \frac{L}{t}\tanh{\left(\frac{xL}{\nu t}\right)}, \\
\label{eq:solBz}
{b_z}_\pm(x,t) &=& \pm b_{z,0} \frac{L}{t}\tanh{\left(\frac{xL}{\nu t}\right)}, 
\ea
where by definition
\be
b_{y,0} = \frac{1}{\sqrt{1+c^2}}, \quad b_{z,0} = \frac{c}{\sqrt{1+c^2}},
\ee
and $c\in \mathbb{R}$ is a constant.
In the limit $\nu \to 0^+$, the solutions (\ref{eq:solU})--(\ref{eq:solBz}) tend to the following inviscid profiles corresponding to discontinuities
\ba 
\label{eq:inviscid_Sol_u}
    u(x,t) &\xrightarrow{\nu\to 0^+}&
    \begin{dcases}
    (x+L)/t, & \mathrm{if} \: x < 0 \\
    (x-L)/t, & \mathrm{if} \: x > 0
    \end{dcases}, \\
\label{eq:inviscid_Sol_by}
    {b_y}_\pm(x,t) &\xrightarrow{\nu\to 0^+}&\pm
    \begin{dcases}
     b_{y,0}L/t, & \mathrm{if} \: x < 0 \\
    -b_{y,0}L/t, & \mathrm{if} \: x > 0
    \end{dcases}, \\
\label{eq:inviscid_Sol_bz}
    {b_z}_\pm(x,t) &\xrightarrow{\nu\to 0^+}&\pm
    \begin{dcases}
     b_{z,0}L/t, & \mathrm{if} \: x < 0 \\
    -b_{z,0}L/t, & \mathrm{if} \: x > 0
    \end{dcases}.
\ea
These are stationary shocks of amplitude $\Delta=2L/t$ for the velocity, $\Delta_y= b_{y,0}\Delta$ for the $y$-component and $\Delta_z=b_{z,0}\Delta$ for the $z$-component of the magnetic field, localized at point $x = 0$ for all time (see Fig.\,\ref{fig:Yanase_Fields}). These inviscid solutions are the MHD version of the Khokhlov sawtooth solution for Burgers' equation \cite{Khokhlov61}. 

It is interesting to note that the analytical solutions (\ref{eq:inviscid_Sol_u})--(\ref{eq:inviscid_Sol_bz}) are fully consistent with the 3D compressible MHD equations since the velocity derivative is constant in space. Indeed, from the Yanase equations completed with the density equation (so far left out of our 1D model), it is possible to find, in the inviscid limit, an exact solution for the mass density, which remains constant (in a given interval) in space but evolves in time as 
\be
\rho(x,t) = \frac{1}{t}[C_+ H(x) + C_- H(-x)] , \label{rhows}
\ee
where $H$ is the Heaviside function and $C_{\pm} \in \mathbb{R}_+^*$. 

A direct numerical simulation of Eqs. (\ref{eq:dudtYanase})--(\ref{eq:dbzdtYanase}) was performed to check if (and how) the analytical solutions (\ref{eq:solU})--(\ref{eq:solBz}) are generated. We use a space resolution of $8192$ grid points and $-L \le x \le L$, with $L=1$. A centered Euler numerical scheme was implemented for both equations with periodic boundary conditions. The initial conditions correspond to plane waves and are $u(x,t=0)=\sin(2\pi x)$, $b_y(x,t=0)=\sin(\pi x)$ and $b_z(x,t=0)=2\sin(\pi x)$. (See \cite{Yanase97} for random initial conditions.)
The timestep is $dt=10^{-6}$ and the viscosity is $\nu = \eta = 1.5 \times 10^{-3}$.
The time evolution of the fields is shown in Fig.\,\ref{fig:Yanase_Fields} from $t=0$ (blue) to $t=1.1$ (red). After an initial phase during which the amplitude of the magnetic field increases locally at the expense of the velocity (with $\max \vert {b_j}_+(x,t>0) \vert > \max \vert {b_j}_+(x,t=0) \vert $ and $\max \vert u(x,t=0) \vert > \max \vert u(x,t>0) \vert$), which can be interpreted as a kind of dynamo effect, the analytical solutions $(u,b_{j\pm})$ are eventually formed at positions $x=0$ and $\pm L$ after the merger of shocks.

\section{\label{sec:meanRates} Mean dissipation rates}
We shall compute the mean rate of energy dissipation. From the analytical solutions (\ref{eq:solU})--(\ref{eq:solBz}) one can find the mean rate of viscous dissipation $\varepsilon \equiv \left\langle \mathcal{D}_{\nu, \nu} (x,t) \right\rangle$,  where $\langle \rangle$ is taken as the space average (\textit{i.e.} $\langle f \rangle = (2L)^{-1}\int_{-L}^{+L} f(x)\mathrm{d}x$). In the limit of small viscosity, a simple calculation gives
\be
\label{dissipnu}
\begin{split}
\lim_{\nu \to 0^{+}} \varepsilon = \lim_{\nu \to 0^{+}} \frac{\nu}{2L} \int_{-L}^{+L} \left\{ \left[ \frac{1}{t} - \frac{L^2}{\nu t^2} \mathrm{sech}^2 \left(\frac{xL}{\nu t} \right) \right]^2 \right.  \\
\left. + \left(b_{y,0}^2 + b_{z,0}^2 \right) \frac{L^4}{t^4 \nu^2} \mathrm{sech}^4 \left(\frac{xL}{\nu t} \right) \right\} \mathrm{d}x. \nonumber
\end{split}
\ee
At the main order, one finds
\be
\label{dissipnu2}
\lim_{\nu \to 0} \varepsilon = \lim_{\nu \to 0}\frac{L^3}{\nu t^4} \int_{-L}^{+L} \mathrm{sech}^4 \left(\frac{xL}{\nu t} \right) \mathrm{d}x = \frac{\Delta^3}{6L} ,
\ee
which is positive and independent of $\nu$. Taking now the inviscid/ideal solutions (\ref{eq:inviscid_Sol_u})--(\ref{eq:inviscid_Sol_bz}), one finds for the variation of total energy  $dE/dt=- \Delta^{3}/(6L)$, which is compatible with expression (\ref{dissipnu2}). However, when $\nu=\eta=0$ the only way to dissipate energy is through the mean inertial dissipation $\left\langle \mathcal{D}_I \right\rangle$. In the inviscid/ideal case, one has (with expressions (\ref{eq:inviscid_Sol_u})--(\ref{eq:inviscid_Sol_bz})), 
\be
\mathcal{D}_I = \lim_{\ell \to 0} \mathcal{D}_I^\ell = \frac{\Delta^3}{3} \delta(x), 
\ee
hence we find the exact relation
\be \label{eq:mean_Di}
\left\langle \mathcal{D}_I \right\rangle = \frac{1}{2L} \int_{-L}^{+L} \frac{\delta(x)}{12} \left[ 1 + 3 \left( b_{y,0}^2 + b_{z,0}^2 \right) \right] \Delta^3 \mathrm{d}x = \frac{\Delta^3}{6L}, \nonumber
\ee
which is equal to $\varepsilon$ in the limit of small viscosity. Therefore, the viscous dissipation can be substituted exactly by the inertial (or anomalous) dissipation in absence of viscosity. In other words, the loss of energy in Yanase's equations is then produced by the loss of regularity of the velocity and magnetic field. These results prove the zeroth law of turbulence in this particular case. 
Note that shocks are often considered as a part of the structures produced by turbulence (for Burgers' equation see \citep{Bec2007}) in addition to fluctuations, as one can see in supersonic hydrodynamic turbulence (see e.g. \cite{F20}). Therefore, the zeroth law of turbulence can be applied to 1D equations too, which are characterized by shock steepening. The particularity of our approach is that the computation of the inertial dissipation does not require a statistical treatment and can be applied to a single nonlinear event.

\begin{figure}
\includegraphics[width=\columnwidth]{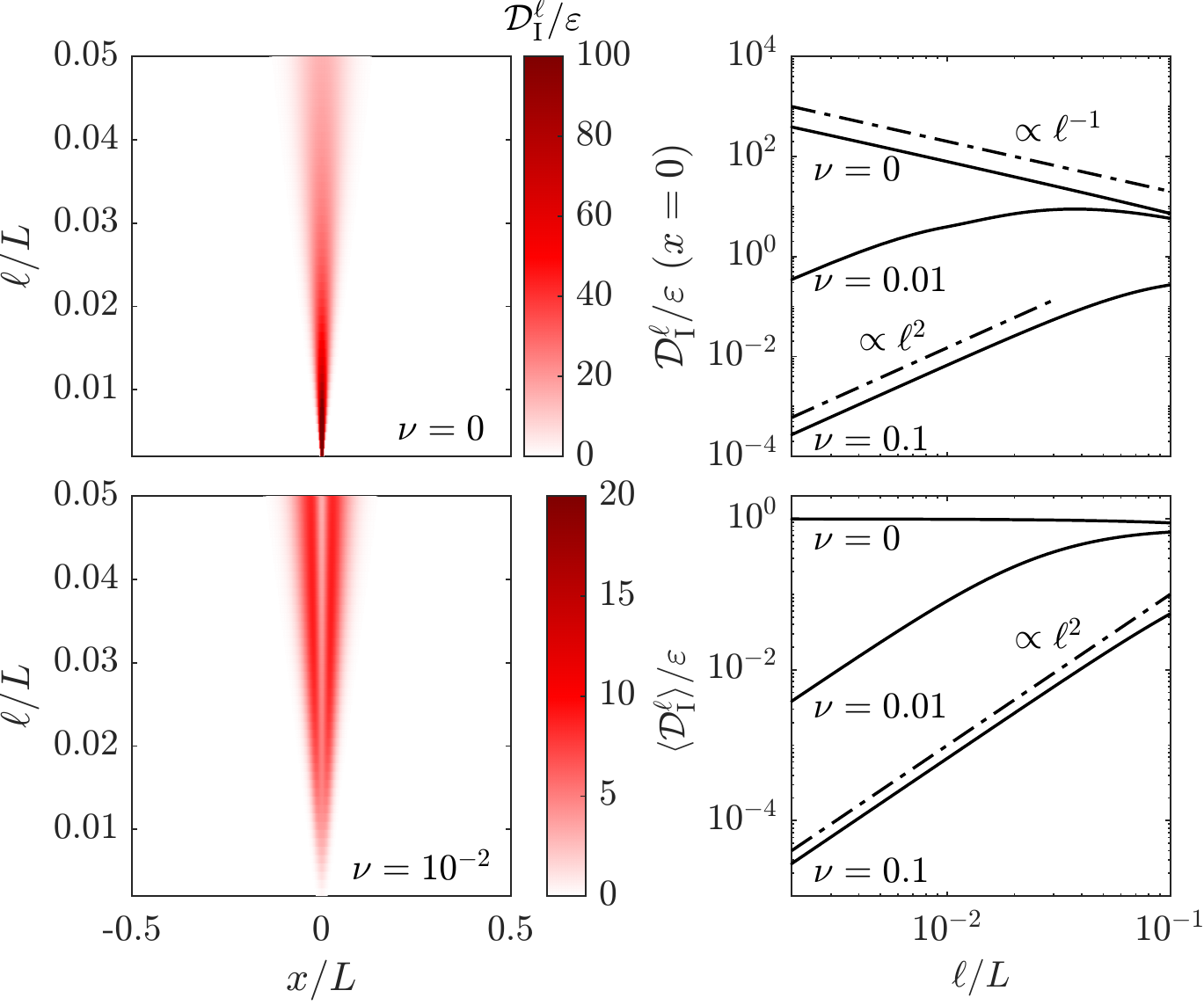}
\caption{Left: space/scale diagram of the normalized local energy transfer $\mathcal{D}_I^\ell / \varepsilon$ for the inviscid (top) and viscous (bottom) analytical solutions of Yanase's equations. The intensity of $\mathcal{D}_I^\ell / \varepsilon$ is given by the colorbar. Right: variation with $\ell$ of the normalized local energy transfer at $x=0$ (top) and averaged over $[-L,+L]$ (bottom), in the inviscid and viscous cases.}
\label{fig:YanaseSliceTheo}
\end{figure}

The discrepancy between the viscous (\ref{eq:solU})--(\ref{eq:solBz}) and the inviscid (\ref{eq:inviscid_Sol_u})--(\ref{eq:inviscid_Sol_bz}) solutions can be investigated numerically with expression (\ref{eq:inertial_dissipation}). The latter is computed with a continuous 1D wavelet transform based on FFT (MATLAB package provided by the toolbox YAWTB) and a normalized Gaussian function is chosen for $\varphi$. Simulations are made on a spatial domain of size $2L=2$ with spatial increment $dx=10^{-3}$. The term $\mathcal{D}_I^\ell$ is computed over the entire simulation domain for $10^2$ values of $\ell \in [1, 10^2]dx$.
Fig. \ref{fig:YanaseSliceTheo} (left) shows the evolution of the inertial dissipation as a function of two parameters: the position in the simulation domain (x-axis) and the width (or scale) of $\varphi$ (y-axis). Both the inviscid (top) and viscous (bottom) solutions lead to a map where the intensity of $\mathcal{D}_I^\ell$ is relatively low far from the shock (localized at $x=0$). It becomes relatively high close to $x=0$ with, however, a difference between the two types of solution: while the inviscid solution continues to increase and forms a true singularity, the viscous one converges to zero at small $\ell$, meaning that it is a quasi-singularity. 
Therefore, this map tells us at first glance what kind of solution leads to a non-zero inertial dissipation. 

A more precise analysis can be done by observing how $\mathcal{D}_I^\ell$ evolves with $\ell$. Fig.\,\ref{fig:YanaseSliceTheo} (top right) shows this (normalized) evolution on the shock for the inviscid case and two viscous cases. We see that the inviscid solution follows a power law in $\ell^{-1}$ at all scales while the viscous solutions follow mainly (at small scale) a power law in $\ell^2$. These behaviors can be understood by simple dimensional arguments. In the inviscid case, $\delta u \sim \delta b \sim \Delta$ and we find $\mathcal{D}_I^\ell \sim \Delta^{3} / \ell$, whereas with the viscosity effect we have $\delta u \sim \delta b \sim \ell$; plugging this into (\ref{eq:inertial_dissipation}) gives $\mathcal{D}_I^\ell \sim \ell^2$. Fig.\,\ref{fig:YanaseSliceTheo} (bottom right) confirms this behavior when the (normalized) mean value $\langle \mathcal{D}_I^\ell \rangle$ is taken. We can conclude that the lower the viscosity, the better the power law $\ell^{-1}$ will be followed by the viscous solution, but one always reaches a scale in $\ell$ below which the viscous solution becomes attractive. 

\begin{figure}
\includegraphics[width=\columnwidth]{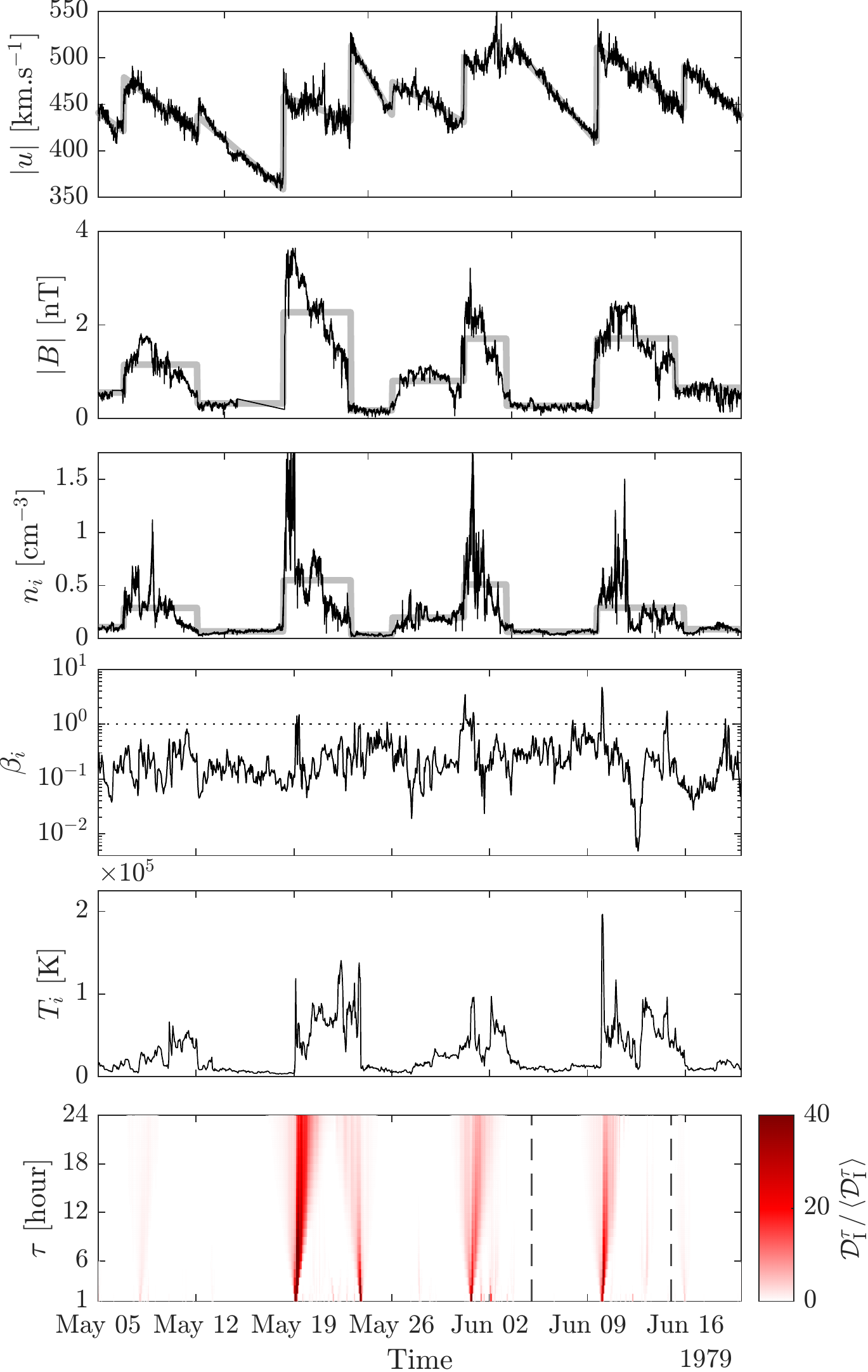}
\caption{From top to bottom: modulus of the proton velocity and magnetic field, the proton density, proton plasma beta, proton temperature and time/scale diagram of the normalized local inertial dissipation (vertical dashed lines delimit the area studied in Fig. \ref{fig:DiMean}). Data measured by Voyager 2 into the solar wind from May 5 to June 20, 1979 at $\sim5.1$\,AU. Theoretical fits (thick grey lines) are superposed to emphasize the sawtooth and crenellated nature of the basic fields.}
\label{fig:Yanase_fig3}
\end{figure}

\section{\label{sec:solarWind} Shocks in the Solar wind}
For $r>10$\,AU cosmic particles are the main source of heating of the solar wind \cite{Pine2020,Pine2020b}, while for $r<2$\,AU turbulent fluctuations are omnipresent. On the other hand, at 2--10\,AU collisionless shocks have been clearly detected by Voyager 1 \& 2 \cite{Burlaga1984,Roberts1987} 
(relatively weak fields fluctuations are also present with possibly a non-trivial dynamics \cite{Malara1999}).
Although these events are known for years and believed to have a strong impact on the local heating \cite{Gazis1982}, to the best of our knowledge, so far no theory
with an exact, concise and directly applicable formula for estimating the heating rate has been proposed.
As we can see on Fig.\ref{fig:Yanase_fig3}, the velocity profile shows shocks on which turbulent fluctuations are superposed.  For the magnetic field and density we also observe large scale structures as well as fluctuations. The study of these fluctuations has  already  been  done  by  \citep{Pine2020} to deduce the turbulent  heating. Here, we only consider the large scale variations of the fields where the fluctuations are filtered.
Our theory may be applied precisely in this region where the radial direction is identified with the x-direction. Then, $u$ is the main component of the velocity:
we have verified in the data that the perpendicular (to the radial direction) components of the velocity are at most equal to $5\%$ (in modulus) of the radial component. Knowing that Parker's spiral angle is close to $80^{o}$ at 5AU, the  transverse (to the radial direction) components of the magnetic field are dominant and $B_x$ can be neglected.
We use the shock profiles (\ref{eq:inviscid_Sol_u})--(\ref{rhows}) which are exact solutions of the 1D compressible MHD equations (Yanase's equations completed by the density equation).
Although limited, it is believed that this 1D MHD approach can provide an estimate of the shock heating in the outer solar wind. The hypothesis that  the dissipation finds its origin in the lack of regularity of the fields is a useful mathematical model for obtaining predictions, but in reality dissipation is physically produced by kinetic effects at sub-MHD scales. Therefore, the heating rate found must be seen as an approximation of the actual dissipation.

In Fig. \ref{fig:Yanase_fig3} an example of such shocks (one can easily verify that these are not contact discontinuities for which only the mass density varies, nor  rotational discontinuities for which $\vert B \vert$ does not vary \citep{G16}) is reported and fitted with a sawtooth velocity (in modulus), a crenellated magnetic field (in modulus) and density. The magnetic field data and plasma moments (density, velocity, temperature) were measured respectively by the Flux Gate Magnetometer (MAG) and the Plasma Spectrometer (PLS) onboard Voyager 2. Data have between 48\,s and 96\,s time resolution \cite{Matthaeus1982}. A moving median window of length 5 points is used to fill small data gaps. The remaining gaps are filled by linear interpolation. 
\begin{figure}
    \centering
    \includegraphics[width=\columnwidth]{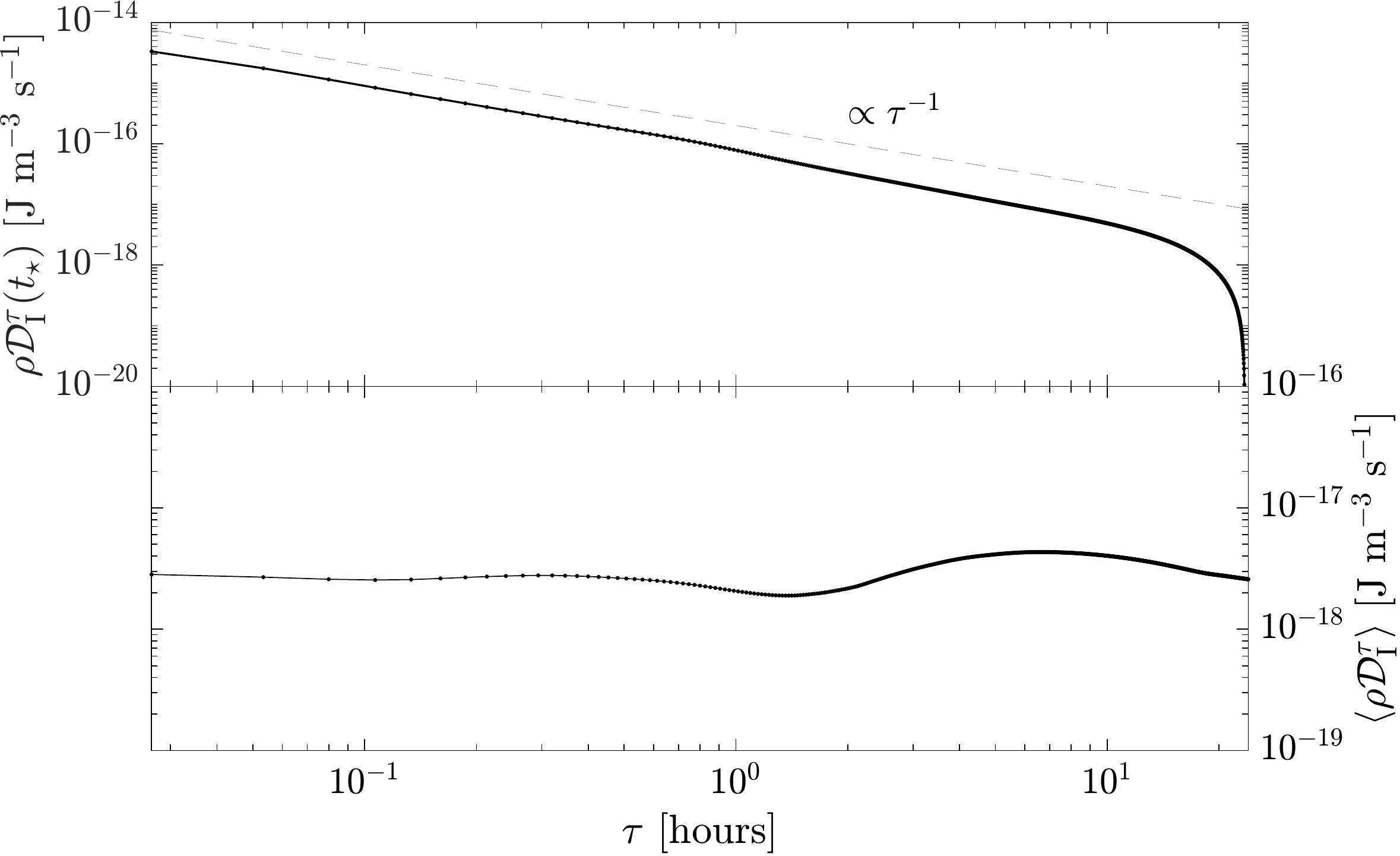}
    \caption{Top: $\rho \mathcal{D}_I^\tau(t_\star)$ as a function of scale $\tau$; note that $t_\star$ corresponds to the shock's location. Bottom: variation with $\tau$ of $\rho \mathcal{D}_I^\tau$ averaged over a time interval of $10$ days centered at the shock position $t_\star$.}
    \label{fig:DiMean}
\end{figure}
After using the Taylor hypothesis ($t=-x/U_{SW}$, with $U_{SW}$ the solar wind velocity), the velocity antishocks can be interpreted as a succession of shocks. 
Therefore, the profiles of the basic fields ($\vert u \vert$, $\vert B \vert$, $n_i$) are roughly compatible with the inviscid solutions (with possibly the absolute value and a normalisation for the magnetic field) discussed above. The 1D MHD model assumes a small plasma beta, a condition satisfied with a (proton) $\beta_i = 2 \mu_0 n_i k_B T_i /B^2$ ($k_B$ the Boltzmann constant) usually smaller than $1$ (see also \cite{Pine2020b}) except on some shock locations (which shows the limitation of our model). The ion temperature reveals very often the presence of large peaks at the discontinuities (e.g. May 19, May 24 or June 10), but it is not systematically the case. A possible explanation is that the energy from the shock may be converted into another form than heating 
(e.g. radiation), hence a possible slight overestimation of the heating. Note that we are in the same situation as for the estimation of the heating rate from the turbulent fluctuations at MHD scales for which it is implicitly assumed that $\varepsilon$ is entirely converted into plasma heating. 
The other smaller peaks observed at any time may be the consequence of a turbulent heating. 
Finally, the time/scale diagram ($\tau$ is used instead of $\ell$) of the normalized local inertial dissipation reveals the position of the main discontinuities which are characterized by a scaling $\mathcal{D}_I^\tau \sim 1/\tau$ (see Fig. \ref{fig:DiMean}). It is actually from this signal we were able to define (most of) the positions of the singularities in the basic fields (thick grey lines). We see that the heating coincides well with peaks in temperature. 

It is straightforward to deduce the heating rate produced by such discontinuities. For the velocity, we find a typical amplitude $\Delta \simeq 10^5$\,m/s and a duration $\tau \simeq 5$ days which is consistent with the integral scale of the $f^{-2}$ spectra measured in this region \cite{Burlaga1987,Burlaga1987b}.
With $U_{SW} \simeq 4.5\times 10^5$m/s, we obtain $L = \tau U_{SW} \simeq 10^{11}$m. The mass density being $n_i \simeq 10^{5}$\,m$^{-3}$, we eventually obtain the mean rate of energy dissipation (or heating rate) 
\be \label{heating}
\rho \left\langle \mathcal{D}_I \right\rangle = \frac{1}{6}\frac{\rho \Delta^3}{L} \sim 10^{-18} {\rm J}\,{\rm m}^{-3}\,{\rm s}^{-1} .
\ee
We can easily verify that the amplitude and duration of the magnetic events (after normalization) are approximately the same, confirming the relevance of our MHD solutions. 

Expression (\ref{heating}) is used to evaluate the local heating produced by a compressible shock localized at $t_\star$, which is the center of the time interval of 10 days (from June 5 to June 15) defined with the two vertical dashed lines in Fig. \ref{fig:Yanase_fig3}. First, the presence of a shock at $t_\star$ is clearly confirmed in Fig. \ref{fig:DiMean} (top) with a clear scaling $\rho \mathcal{D}_I^\tau \sim 1/\tau$. Then, the local heating can be directly deduced from the inertial dissipation averaged over this time interval of 10 days centered around $t_\star$ (bottom). Within the limit of small $\tau$, a value $\left\langle \rho \mathcal{D}_I^\tau \right\rangle \simeq 3 \times 10^{-18}$J\,m$^{-3}$\,s$^{-1}$ is obtained, which confirms the previous evaluation. 
Note that the overall behavior found here should be compared to Fig. \ref{fig:YanaseSliceTheo} (right). The heating rate found is smaller than that commonly obtained at $1$\,AU from turbulent fluctuations, with 
$\rho \varepsilon > 10^{-17}$J\,m$^{-3}$\,s$^{-1}$ \cite{sorriso2007,Vasquez07,macbride08,Marino08,Carbone09,Stawarz09,coburn15,Banerjee2016,Hadid2018,Andres2019}. 
However, the value (\ref{heating}) is more than one order of magnitude larger than the values recently reported  from the turbulent fluctuations at the same heliocentric distance \cite{Pine2020}. This suggests that in the outer solar wind, discontinuities, if they exist, can be the main source of local heating.

\section{\label{sec:conclusion} Conclusion}
Understanding the mechanisms of dissipation in collisionless plasmas remains a subject of great importance in space physics, astrophysics, and laboratory plasmas. Although plasma heating involves kinetic effects at sub-ion scales, it has been increasingly recognized in recent years that the exact laws of turbulence in MHD provide a means of estimating the mean rate of heating through the measure of the mean rate of energy transfer \cite{sorriso2007,Vasquez07,macbride08,Marino08,Carbone09,Stawarz09,coburn15,Banerjee2016,Hadid2018,Andres2019}. The basic assumption is that the MHD energy transfer is mainly converted into small-scale heating. For space plasmas, this method has been used in homogeneous regions (solar wind and Earth's magnetosheath) where turbulent fluctuations are dominant. However, in presence of collisionless shocks, the assumption of statistical homogeneity is broken and the exact laws of turbulence become unusable \cite{G18}.

In this article, we propose a new method to estimate the mean rate of heating produced by shocks. 
Note that the problem of shock heating has already been studied in the context of coronal heating in a superradiallly open flux tube, but for which the 1D MHD equations are different and no analytical prediction have been made \citep{Suzuki2005}.
As with the exact laws of turbulence, the estimate made here from MHD must be seen as an approximation of the actual heating. In the case of collisional plasmas like the solar wind, the non-regularity assumed for the basic fields is certainly broken at sub-MHD scales where in particular kinetic effects drive the plasma heating. 
An application to the solar wind of the 1D MHD solutions, which are by nature intrinsically limited, reveals that this heating is much higher than the values obtained from turbulent fluctuations at the same heliocentric distance \cite{Pine2020}, suggesting that collisionless shocks can be a dominant source of heating in the outer solar wind. Given the importance of collisionless shocks in astrophysics and laboratory plasmas \cite{Miceli2019}, it is believed that our study may be useful in a number of other situations.

The zeroth law of turbulence is one of the oldest conjecture in turbulence that is still unproven in general. The system considered in this article provides an example where this law can be proven. This reinforces Onsager's conjecture \cite{Onsager49} who assumed that the lack of smoothness could be the cause of the energy dissipation in the limit of infinite Reynolds numbers. 
Our result is encouraging but the gap for a complete demonstration at the level of 3D Navier-Stokes equations remains immense.

\acknowledgements
V.D. acknowledges B. Dubrulle for helpful discussion.

\bibliography{biblio}

\providecommand{\noopsort}[1]{}\providecommand{\singleletter}[1]{#1}%
\begin{thebibliography}{53}%
\makeatletter
\providecommand \@ifxundefined [1]{%
 \@ifx{#1\undefined}
}%
\providecommand \@ifnum [1]{%
 \ifnum #1\expandafter \@firstoftwo
 \else \expandafter \@secondoftwo
 \fi
}%
\providecommand \@ifx [1]{%
 \ifx #1\expandafter \@firstoftwo
 \else \expandafter \@secondoftwo
 \fi
}%
\providecommand \natexlab [1]{#1}%
\providecommand \enquote  [1]{``#1''}%
\providecommand \bibnamefont  [1]{#1}%
\providecommand \bibfnamefont [1]{#1}%
\providecommand \citenamefont [1]{#1}%
\providecommand \href@noop [0]{\@secondoftwo}%
\providecommand \href [0]{\begingroup \@sanitize@url \@href}%
\providecommand \@href[1]{\@@startlink{#1}\@@href}%
\providecommand \@@href[1]{\endgroup#1\@@endlink}%
\providecommand \@sanitize@url [0]{\catcode `\\12\catcode `\$12\catcode
  `\&12\catcode `\#12\catcode `\^12\catcode `\_12\catcode `\%12\relax}%
\providecommand \@@startlink[1]{}%
\providecommand \@@endlink[0]{}%
\providecommand \url  [0]{\begingroup\@sanitize@url \@url }%
\providecommand \@url [1]{\endgroup\@href {#1}{\urlprefix }}%
\providecommand \urlprefix  [0]{URL }%
\providecommand \Eprint [0]{\href }%
\providecommand \doibase [0]{http://dx.doi.org/}%
\providecommand \selectlanguage [0]{\@gobble}%
\providecommand \bibinfo  [0]{\@secondoftwo}%
\providecommand \bibfield  [0]{\@secondoftwo}%
\providecommand \translation [1]{[#1]}%
\providecommand \BibitemOpen [0]{}%
\providecommand \bibitemStop [0]{}%
\providecommand \bibitemNoStop [0]{.\EOS\space}%
\providecommand \EOS [0]{\spacefactor3000\relax}%
\providecommand \BibitemShut  [1]{\csname bibitem#1\endcsname}%
\let\auto@bib@innerbib\@empty
\bibitem [{\citenamefont {Onsager}(1949)}]{Onsager49}%
  \BibitemOpen
  \bibfield  {author} {\bibinfo {author} {\bibfnamefont {L.}~\bibnamefont
  {Onsager}},\ }\href {\doibase 10.1007/BF02780991} {\bibfield  {journal}
  {\bibinfo  {journal} {Il Nuovo Cimento}\ }\textbf {\bibinfo {volume} {6}}
  (\bibinfo {year} {1949}),\ 10.1007/BF02780991}\BibitemShut {NoStop}%
\bibitem [{\citenamefont {Frisch}(1995)}]{Frisch95}%
  \BibitemOpen
  \bibfield  {author} {\bibinfo {author} {\bibfnamefont {U.}~\bibnamefont
  {Frisch}},\ }\href {\doibase 10.1017/CBO9781139170666} {\emph {\bibinfo
  {title} {Turbulence: The Legacy of A. N. Kolmogorov}}}\ (\bibinfo
  {publisher} {Cambridge University Press},\ \bibinfo {year}
  {1995})\BibitemShut {NoStop}%
\bibitem [{\citenamefont {Eyink}(1994)}]{Eyink94}%
  \BibitemOpen
  \bibfield  {author} {\bibinfo {author} {\bibfnamefont {G.~L.}\ \bibnamefont
  {Eyink}},\ }\href {\doibase https://doi.org/10.1016/0167-2789(94)90117-1}
  {\bibfield  {journal} {\bibinfo  {journal} {Physica D: Nonlinear Phenomena}\
  }\textbf {\bibinfo {volume} {78}},\ \bibinfo {pages} {222 } (\bibinfo {year}
  {1994})}\BibitemShut {NoStop}%
\bibitem [{\citenamefont {Saint-Michel}\ \emph {et~al.}(2014)\citenamefont
  {Saint-Michel}, \citenamefont {Herbert}, \citenamefont {Salort},
  \citenamefont {Baudet}, \citenamefont {Bon~Mardion}, \citenamefont {Bonnay},
  \citenamefont {Bourgoin}, \citenamefont {Castaing}, \citenamefont
  {Chevillard}, \citenamefont {Daviaud}, \citenamefont {Diribarne},
  \citenamefont {Dubrulle}, \citenamefont {Gagne}, \citenamefont {Gibert},
  \citenamefont {Girard}, \citenamefont {Hébral}, \citenamefont {Lehner},\
  and\ \citenamefont {Rousset}}]{SaintMichel2014}%
  \BibitemOpen
  \bibfield  {author} {\bibinfo {author} {\bibfnamefont {B.}~\bibnamefont
  {Saint-Michel}}, \bibinfo {author} {\bibfnamefont {E.}~\bibnamefont
  {Herbert}}, \bibinfo {author} {\bibfnamefont {J.}~\bibnamefont {Salort}},
  \bibinfo {author} {\bibfnamefont {C.}~\bibnamefont {Baudet}}, \bibinfo
  {author} {\bibfnamefont {M.}~\bibnamefont {Bon~Mardion}}, \bibinfo {author}
  {\bibfnamefont {P.}~\bibnamefont {Bonnay}}, \bibinfo {author} {\bibfnamefont
  {M.}~\bibnamefont {Bourgoin}}, \bibinfo {author} {\bibfnamefont
  {B.}~\bibnamefont {Castaing}}, \bibinfo {author} {\bibfnamefont
  {L.}~\bibnamefont {Chevillard}}, \bibinfo {author} {\bibfnamefont
  {F.}~\bibnamefont {Daviaud}}, \bibinfo {author} {\bibfnamefont
  {P.}~\bibnamefont {Diribarne}}, \bibinfo {author} {\bibfnamefont
  {B.}~\bibnamefont {Dubrulle}}, \bibinfo {author} {\bibfnamefont
  {Y.}~\bibnamefont {Gagne}}, \bibinfo {author} {\bibfnamefont
  {M.}~\bibnamefont {Gibert}}, \bibinfo {author} {\bibfnamefont
  {A.}~\bibnamefont {Girard}}, \bibinfo {author} {\bibfnamefont
  {B.}~\bibnamefont {Hébral}}, \bibinfo {author} {\bibfnamefont
  {T.}~\bibnamefont {Lehner}}, \ and\ \bibinfo {author} {\bibfnamefont
  {B.}~\bibnamefont {Rousset}},\ }\href {\doibase 10.1063/1.4904378} {\bibfield
   {journal} {\bibinfo  {journal} {Physics of Fluids}\ }\textbf {\bibinfo
  {volume} {26}},\ \bibinfo {pages} {125109} (\bibinfo {year}
  {2014})}\BibitemShut {NoStop}%
\bibitem [{\citenamefont {Galtier}(2016)}]{G16}%
  \BibitemOpen
  \bibfield  {author} {\bibinfo {author} {\bibfnamefont {S.}~\bibnamefont
  {Galtier}},\ }\href {\doibase 10.1017/CBO9781316665961} {\emph {\bibinfo
  {title} {Introduction to modern magnetohydrodynamics}}}\ (\bibinfo
  {publisher} {Cambridge University Press},\ \bibinfo {year} {2016})\ p.\
  \bibinfo {pages} {288}\BibitemShut {NoStop}%
\bibitem [{\citenamefont {Mininni}\ and\ \citenamefont
  {Pouquet}(2009)}]{Mininni09}%
  \BibitemOpen
  \bibfield  {author} {\bibinfo {author} {\bibfnamefont {P.~D.}\ \bibnamefont
  {Mininni}}\ and\ \bibinfo {author} {\bibfnamefont {A.}~\bibnamefont
  {Pouquet}},\ }\href {\doibase 10.1103/PhysRevE.80.025401} {\bibfield
  {journal} {\bibinfo  {journal} {Phys. Rev. E}\ }\textbf {\bibinfo {volume}
  {80}},\ \bibinfo {pages} {025401(R)} (\bibinfo {year} {2009})}\BibitemShut
  {NoStop}%
\bibitem [{\citenamefont {Bandyopadhyay}\ \emph {et~al.}(2018)\citenamefont
  {Bandyopadhyay}, \citenamefont {Oughton}, \citenamefont {Wan}, \citenamefont
  {Matthaeus}, \citenamefont {Chhiber},\ and\ \citenamefont
  {Parashar}}]{Bandyopadhyay18}%
  \BibitemOpen
  \bibfield  {author} {\bibinfo {author} {\bibfnamefont {R.}~\bibnamefont
  {Bandyopadhyay}}, \bibinfo {author} {\bibfnamefont {S.}~\bibnamefont
  {Oughton}}, \bibinfo {author} {\bibfnamefont {M.}~\bibnamefont {Wan}},
  \bibinfo {author} {\bibfnamefont {W.~H.}\ \bibnamefont {Matthaeus}}, \bibinfo
  {author} {\bibfnamefont {R.}~\bibnamefont {Chhiber}}, \ and\ \bibinfo
  {author} {\bibfnamefont {T.~N.}\ \bibnamefont {Parashar}},\ }\href {\doibase
  10.1103/PhysRevX.8.041052} {\bibfield  {journal} {\bibinfo  {journal} {Phys.
  Rev. X}\ }\textbf {\bibinfo {volume} {8}},\ \bibinfo {pages} {041052}
  (\bibinfo {year} {2018})}\BibitemShut {NoStop}%
\bibitem [{\citenamefont {Kolmogorov}(1941)}]{K41}%
  \BibitemOpen
  \bibfield  {author} {\bibinfo {author} {\bibfnamefont {A.~N.}\ \bibnamefont
  {Kolmogorov}},\ }\href@noop {} {\bibfield  {journal} {\bibinfo  {journal}
  {Dokl Akad Nauk SSSR}\ }\textbf {\bibinfo {volume} {32}},\ \bibinfo {pages}
  {16} (\bibinfo {year} {1941})}\BibitemShut {NoStop}%
\bibitem [{\citenamefont {Monin}(1994)}]{Monin59}%
  \BibitemOpen
  \bibfield  {author} {\bibinfo {author} {\bibfnamefont {A.~S.}\ \bibnamefont
  {Monin}},\ }\href@noop {} {\bibfield  {journal} {\bibinfo  {journal} {Doklady
  Akademii Nauk SSSR}\ }\textbf {\bibinfo {volume} {125}} (\bibinfo {year}
  {1994})}\BibitemShut {NoStop}%
\bibitem [{\citenamefont {Antonia}\ \emph {et~al.}(1997)\citenamefont
  {Antonia}, \citenamefont {Ould-Rouis}, \citenamefont {Anselmet},\ and\
  \citenamefont {Zhu}}]{Antonia97}%
  \BibitemOpen
  \bibfield  {author} {\bibinfo {author} {\bibfnamefont {R.~A.}\ \bibnamefont
  {Antonia}}, \bibinfo {author} {\bibfnamefont {M.}~\bibnamefont {Ould-Rouis}},
  \bibinfo {author} {\bibfnamefont {F.}~\bibnamefont {Anselmet}}, \ and\
  \bibinfo {author} {\bibfnamefont {Y.}~\bibnamefont {Zhu}},\ }\href {\doibase
  10.1017/S0022112096004090} {\bibfield  {journal} {\bibinfo  {journal}
  {Journal of Fluid Mechanics}\ }\textbf {\bibinfo {volume} {332}},\ \bibinfo
  {pages} {395–409} (\bibinfo {year} {1997})}\BibitemShut {NoStop}%
\bibitem [{\citenamefont {{Lindborg}}(1999)}]{Lindborg1999}%
  \BibitemOpen
  \bibfield  {author} {\bibinfo {author} {\bibfnamefont {E.}~\bibnamefont
  {{Lindborg}}},\ }\href {\doibase 10.1017/S0022112099004851} {\bibfield
  {journal} {\bibinfo  {journal} {J. Fluid Mech.}\ }\textbf {\bibinfo {volume}
  {388}},\ \bibinfo {pages} {259} (\bibinfo {year} {1999})}\BibitemShut
  {NoStop}%
\bibitem [{\citenamefont {Galtier}\ and\ \citenamefont {Banerjee}(2011)}]{G11}%
  \BibitemOpen
  \bibfield  {author} {\bibinfo {author} {\bibfnamefont {S.}~\bibnamefont
  {Galtier}}\ and\ \bibinfo {author} {\bibfnamefont {S.}~\bibnamefont
  {Banerjee}},\ }\href {\doibase 10.1103/PhysRevLett.107.134501} {\bibfield
  {journal} {\bibinfo  {journal} {Phys. Rev. Lett.}\ }\textbf {\bibinfo
  {volume} {107}},\ \bibinfo {pages} {134501} (\bibinfo {year}
  {2011})}\BibitemShut {NoStop}%
\bibitem [{\citenamefont {Ferrand}\ \emph {et~al.}(2020)\citenamefont
  {Ferrand}, \citenamefont {Galtier}, \citenamefont {Sahraoui},\ and\
  \citenamefont {Federrath}}]{F20}%
  \BibitemOpen
  \bibfield  {author} {\bibinfo {author} {\bibfnamefont {R.}~\bibnamefont
  {Ferrand}}, \bibinfo {author} {\bibfnamefont {S.}~\bibnamefont {Galtier}},
  \bibinfo {author} {\bibfnamefont {F.}~\bibnamefont {Sahraoui}}, \ and\
  \bibinfo {author} {\bibfnamefont {C.}~\bibnamefont {Federrath}},\ }\href
  {\doibase 10.3847/1538-4357/abb76e} {\bibfield  {journal} {\bibinfo
  {journal} {The Astrophysical Journal}\ }\textbf {\bibinfo {volume} {904}},\
  \bibinfo {pages} {160} (\bibinfo {year} {2020})}\BibitemShut {NoStop}%
\bibitem [{\citenamefont {Politano}\ and\ \citenamefont
  {Pouquet}(1998)}]{PP98a}%
  \BibitemOpen
  \bibfield  {author} {\bibinfo {author} {\bibfnamefont {H.}~\bibnamefont
  {Politano}}\ and\ \bibinfo {author} {\bibfnamefont {A.}~\bibnamefont
  {Pouquet}},\ }\href {\doibase 10.1103/PhysRevE.57.R21} {\bibfield  {journal}
  {\bibinfo  {journal} {Phys. Rev. E}\ }\textbf {\bibinfo {volume} {57}},\
  \bibinfo {pages} {R21} (\bibinfo {year} {1998})}\BibitemShut {NoStop}%
\bibitem [{\citenamefont {Banerjee}\ and\ \citenamefont
  {Galtier}(2013)}]{BG13}%
  \BibitemOpen
  \bibfield  {author} {\bibinfo {author} {\bibfnamefont {S.}~\bibnamefont
  {Banerjee}}\ and\ \bibinfo {author} {\bibfnamefont {S.}~\bibnamefont
  {Galtier}},\ }\href {\doibase 10.1103/PhysRevE.87.013019} {\bibfield
  {journal} {\bibinfo  {journal} {Phys. Rev. E}\ }\textbf {\bibinfo {volume}
  {87}},\ \bibinfo {pages} {013019} (\bibinfo {year} {2013})}\BibitemShut
  {NoStop}%
\bibitem [{\citenamefont {{Wan}}\ \emph {et~al.}(2009)\citenamefont {{Wan}},
  \citenamefont {{Servidio}}, \citenamefont {{Oughton}},\ and\ \citenamefont
  {{Matthaeus}}}]{Wan2009}%
  \BibitemOpen
  \bibfield  {author} {\bibinfo {author} {\bibfnamefont {M.}~\bibnamefont
  {{Wan}}}, \bibinfo {author} {\bibfnamefont {S.}~\bibnamefont {{Servidio}}},
  \bibinfo {author} {\bibfnamefont {S.}~\bibnamefont {{Oughton}}}, \ and\
  \bibinfo {author} {\bibfnamefont {W.~H.}\ \bibnamefont {{Matthaeus}}},\
  }\href {\doibase 10.1063/1.3240333} {\bibfield  {journal} {\bibinfo
  {journal} {Phys. Plasmas}\ }\textbf {\bibinfo {volume} {16}},\ \bibinfo
  {pages} {090703} (\bibinfo {year} {2009})}\BibitemShut {NoStop}%
\bibitem [{\citenamefont {Galtier}(2008)}]{G08}%
  \BibitemOpen
  \bibfield  {author} {\bibinfo {author} {\bibfnamefont {S.}~\bibnamefont
  {Galtier}},\ }\href {\doibase 10.1103/PhysRevE.77.015302} {\bibfield
  {journal} {\bibinfo  {journal} {Phys. Rev. E}\ }\textbf {\bibinfo {volume}
  {77}},\ \bibinfo {pages} {015302(R)} (\bibinfo {year} {2008})}\BibitemShut
  {NoStop}%
\bibitem [{\citenamefont {Andr\'es}\ \emph {et~al.}(2018)\citenamefont
  {Andr\'es}, \citenamefont {Galtier},\ and\ \citenamefont
  {Sahraoui}}]{Andres18}%
  \BibitemOpen
  \bibfield  {author} {\bibinfo {author} {\bibfnamefont {N.}~\bibnamefont
  {Andr\'es}}, \bibinfo {author} {\bibfnamefont {S.}~\bibnamefont {Galtier}}, \
  and\ \bibinfo {author} {\bibfnamefont {F.}~\bibnamefont {Sahraoui}},\ }\href
  {\doibase 10.1103/PhysRevE.97.013204} {\bibfield  {journal} {\bibinfo
  {journal} {Phys. Rev. E}\ }\textbf {\bibinfo {volume} {97}},\ \bibinfo
  {pages} {013204} (\bibinfo {year} {2018})}\BibitemShut {NoStop}%
\bibitem [{\citenamefont {Ferrand}\ \emph {et~al.}(2019)\citenamefont
  {Ferrand}, \citenamefont {Galtier}, \citenamefont {Sahraoui}, \citenamefont
  {Meyrand}, \citenamefont {Andr{\'{e}}s},\ and\ \citenamefont
  {Banerjee}}]{F19}%
  \BibitemOpen
  \bibfield  {author} {\bibinfo {author} {\bibfnamefont {R.}~\bibnamefont
  {Ferrand}}, \bibinfo {author} {\bibfnamefont {S.}~\bibnamefont {Galtier}},
  \bibinfo {author} {\bibfnamefont {F.}~\bibnamefont {Sahraoui}}, \bibinfo
  {author} {\bibfnamefont {R.}~\bibnamefont {Meyrand}}, \bibinfo {author}
  {\bibfnamefont {N.}~\bibnamefont {Andr{\'{e}}s}}, \ and\ \bibinfo {author}
  {\bibfnamefont {S.}~\bibnamefont {Banerjee}},\ }\href {\doibase
  10.3847/1538-4357/ab2be9} {\bibfield  {journal} {\bibinfo  {journal} {The
  Astrophysical Journal}\ }\textbf {\bibinfo {volume} {881}},\ \bibinfo {pages}
  {50} (\bibinfo {year} {2019})}\BibitemShut {NoStop}%
\bibitem [{\citenamefont {{Gazis}}\ and\ \citenamefont
  {{Lazarus}}(1982)}]{Gazis1982}%
  \BibitemOpen
  \bibfield  {author} {\bibinfo {author} {\bibfnamefont {P.~R.}\ \bibnamefont
  {{Gazis}}}\ and\ \bibinfo {author} {\bibfnamefont {A.~J.}\ \bibnamefont
  {{Lazarus}}},\ }\href {\doibase https://doi.org/10.1029/GL009i004p00431}
  {\bibfield  {journal} {\bibinfo  {journal} {Geophysical Research Letters}\
  }\textbf {\bibinfo {volume} {9}},\ \bibinfo {pages} {431} (\bibinfo {year}
  {1982})}\BibitemShut {NoStop}%
\bibitem [{\citenamefont {{Richardson}}\ \emph {et~al.}(1995)\citenamefont
  {{Richardson}}, \citenamefont {{Paularena}}, \citenamefont {{Lazarus}},\ and\
  \citenamefont {{Belcher}}}]{Richardson1995}%
  \BibitemOpen
  \bibfield  {author} {\bibinfo {author} {\bibfnamefont {J.~D.}\ \bibnamefont
  {{Richardson}}}, \bibinfo {author} {\bibfnamefont {K.~I.}\ \bibnamefont
  {{Paularena}}}, \bibinfo {author} {\bibfnamefont {A.~J.}\ \bibnamefont
  {{Lazarus}}}, \ and\ \bibinfo {author} {\bibfnamefont {J.~W.}\ \bibnamefont
  {{Belcher}}},\ }\href {\doibase https://doi.org/10.1029/95GL01421} {\bibfield
   {journal} {\bibinfo  {journal} {Geophysical Research Letters}\ }\textbf
  {\bibinfo {volume} {22}},\ \bibinfo {pages} {1469} (\bibinfo {year}
  {1995})}\BibitemShut {NoStop}%
\bibitem [{\citenamefont {Matthaeus}\ \emph {et~al.}(1999)\citenamefont
  {Matthaeus}, \citenamefont {Zank}, \citenamefont {Smith},\ and\ \citenamefont
  {Oughton}}]{Matthaeus99}%
  \BibitemOpen
  \bibfield  {author} {\bibinfo {author} {\bibfnamefont {W.~H.}\ \bibnamefont
  {Matthaeus}}, \bibinfo {author} {\bibfnamefont {G.~P.}\ \bibnamefont {Zank}},
  \bibinfo {author} {\bibfnamefont {C.~W.}\ \bibnamefont {Smith}}, \ and\
  \bibinfo {author} {\bibfnamefont {S.}~\bibnamefont {Oughton}},\ }\href
  {\doibase 10.1103/PhysRevLett.82.3444} {\bibfield  {journal} {\bibinfo
  {journal} {Phys. Rev. Lett.}\ }\textbf {\bibinfo {volume} {82}},\ \bibinfo
  {pages} {3444} (\bibinfo {year} {1999})}\BibitemShut {NoStop}%
\bibitem [{\citenamefont {Sorriso-Valvo}\ \emph {et~al.}(2007)\citenamefont
  {Sorriso-Valvo}, \citenamefont {Marino}, \citenamefont {Carbone},
  \citenamefont {Noullez}, \citenamefont {Lepreti}, \citenamefont {Veltri},
  \citenamefont {Bruno}, \citenamefont {Bavassano},\ and\ \citenamefont
  {Pietropaolo}}]{sorriso2007}%
  \BibitemOpen
  \bibfield  {author} {\bibinfo {author} {\bibfnamefont {L.}~\bibnamefont
  {Sorriso-Valvo}}, \bibinfo {author} {\bibfnamefont {R.}~\bibnamefont
  {Marino}}, \bibinfo {author} {\bibfnamefont {V.}~\bibnamefont {Carbone}},
  \bibinfo {author} {\bibfnamefont {A.}~\bibnamefont {Noullez}}, \bibinfo
  {author} {\bibfnamefont {F.}~\bibnamefont {Lepreti}}, \bibinfo {author}
  {\bibfnamefont {P.}~\bibnamefont {Veltri}}, \bibinfo {author} {\bibfnamefont
  {R.}~\bibnamefont {Bruno}}, \bibinfo {author} {\bibfnamefont
  {B.}~\bibnamefont {Bavassano}}, \ and\ \bibinfo {author} {\bibfnamefont
  {E.}~\bibnamefont {Pietropaolo}},\ }\href {\doibase
  10.1103/PhysRevLett.99.115001} {\bibfield  {journal} {\bibinfo  {journal}
  {Phys. Rev. Lett.}\ }\textbf {\bibinfo {volume} {99}},\ \bibinfo {pages}
  {115001} (\bibinfo {year} {2007})}\BibitemShut {NoStop}%
\bibitem [{\citenamefont {Vasquez}\ \emph {et~al.}(2007)\citenamefont
  {Vasquez}, \citenamefont {Smith}, \citenamefont {Hamilton}, \citenamefont
  {MacBride},\ and\ \citenamefont {Leamon}}]{Vasquez07}%
  \BibitemOpen
  \bibfield  {author} {\bibinfo {author} {\bibfnamefont {B.~J.}\ \bibnamefont
  {Vasquez}}, \bibinfo {author} {\bibfnamefont {C.~W.}\ \bibnamefont {Smith}},
  \bibinfo {author} {\bibfnamefont {K.}~\bibnamefont {Hamilton}}, \bibinfo
  {author} {\bibfnamefont {B.~T.}\ \bibnamefont {MacBride}}, \ and\ \bibinfo
  {author} {\bibfnamefont {R.~J.}\ \bibnamefont {Leamon}},\ }\href {\doibase
  https://doi.org/10.1029/2007JA012305} {\bibfield  {journal} {\bibinfo
  {journal} {Journal of Geophysical Research: Space Physics}\ }\textbf
  {\bibinfo {volume} {112}} (\bibinfo {year} {2007}),\
  https://doi.org/10.1029/2007JA012305}\BibitemShut {NoStop}%
\bibitem [{\citenamefont {MacBride}\ \emph {et~al.}(2008)\citenamefont
  {MacBride}, \citenamefont {Smith},\ and\ \citenamefont
  {Forman}}]{macbride08}%
  \BibitemOpen
  \bibfield  {author} {\bibinfo {author} {\bibfnamefont {B.~T.}\ \bibnamefont
  {MacBride}}, \bibinfo {author} {\bibfnamefont {C.~W.}\ \bibnamefont {Smith}},
  \ and\ \bibinfo {author} {\bibfnamefont {M.~A.}\ \bibnamefont {Forman}},\
  }\href {\doibase 10.1086/529575} {\bibfield  {journal} {\bibinfo  {journal}
  {The Astrophysical Journal}\ }\textbf {\bibinfo {volume} {679}},\ \bibinfo
  {pages} {1644} (\bibinfo {year} {2008})}\BibitemShut {NoStop}%
\bibitem [{\citenamefont {Marino}\ \emph {et~al.}(2008)\citenamefont {Marino},
  \citenamefont {Sorriso-Valvo}, \citenamefont {Carbone}, \citenamefont
  {Noullez}, \citenamefont {Bruno},\ and\ \citenamefont
  {Bavassano}}]{Marino08}%
  \BibitemOpen
  \bibfield  {author} {\bibinfo {author} {\bibfnamefont {R.}~\bibnamefont
  {Marino}}, \bibinfo {author} {\bibfnamefont {L.}~\bibnamefont
  {Sorriso-Valvo}}, \bibinfo {author} {\bibfnamefont {V.}~\bibnamefont
  {Carbone}}, \bibinfo {author} {\bibfnamefont {A.}~\bibnamefont {Noullez}},
  \bibinfo {author} {\bibfnamefont {R.}~\bibnamefont {Bruno}}, \ and\ \bibinfo
  {author} {\bibfnamefont {B.}~\bibnamefont {Bavassano}},\ }\href {\doibase
  10.1086/587957} {\bibfield  {journal} {\bibinfo  {journal} {The Astrophysical
  Journal}\ }\textbf {\bibinfo {volume} {677}},\ \bibinfo {pages} {L71}
  (\bibinfo {year} {2008})}\BibitemShut {NoStop}%
\bibitem [{\citenamefont {Carbone}\ \emph {et~al.}(2009)\citenamefont
  {Carbone}, \citenamefont {Marino}, \citenamefont {Sorriso-Valvo},
  \citenamefont {Noullez},\ and\ \citenamefont {Bruno}}]{Carbone09}%
  \BibitemOpen
  \bibfield  {author} {\bibinfo {author} {\bibfnamefont {V.}~\bibnamefont
  {Carbone}}, \bibinfo {author} {\bibfnamefont {R.}~\bibnamefont {Marino}},
  \bibinfo {author} {\bibfnamefont {L.}~\bibnamefont {Sorriso-Valvo}}, \bibinfo
  {author} {\bibfnamefont {A.}~\bibnamefont {Noullez}}, \ and\ \bibinfo
  {author} {\bibfnamefont {R.}~\bibnamefont {Bruno}},\ }\href {\doibase
  10.1103/PhysRevLett.103.061102} {\bibfield  {journal} {\bibinfo  {journal}
  {Phys. Rev. Lett.}\ }\textbf {\bibinfo {volume} {103}},\ \bibinfo {pages}
  {061102} (\bibinfo {year} {2009})}\BibitemShut {NoStop}%
\bibitem [{\citenamefont {Stawarz}\ \emph {et~al.}(2009)\citenamefont
  {Stawarz}, \citenamefont {Smith}, \citenamefont {Vasquez}, \citenamefont
  {Forman},\ and\ \citenamefont {MacBride}}]{Stawarz09}%
  \BibitemOpen
  \bibfield  {author} {\bibinfo {author} {\bibfnamefont {J.~E.}\ \bibnamefont
  {Stawarz}}, \bibinfo {author} {\bibfnamefont {C.~W.}\ \bibnamefont {Smith}},
  \bibinfo {author} {\bibfnamefont {B.~J.}\ \bibnamefont {Vasquez}}, \bibinfo
  {author} {\bibfnamefont {M.~A.}\ \bibnamefont {Forman}}, \ and\ \bibinfo
  {author} {\bibfnamefont {B.~T.}\ \bibnamefont {MacBride}},\ }\href {\doibase
  10.1088/0004-637x/697/2/1119} {\bibfield  {journal} {\bibinfo  {journal} {The
  Astrophysical Journal}\ }\textbf {\bibinfo {volume} {697}},\ \bibinfo {pages}
  {1119} (\bibinfo {year} {2009})}\BibitemShut {NoStop}%
\bibitem [{\citenamefont {Coburn}\ \emph {et~al.}(2015)\citenamefont {Coburn},
  \citenamefont {Forman}, \citenamefont {Smith}, \citenamefont {Vasquez},\ and\
  \citenamefont {Stawarz}}]{coburn15}%
  \BibitemOpen
  \bibfield  {author} {\bibinfo {author} {\bibfnamefont {J.~T.}\ \bibnamefont
  {Coburn}}, \bibinfo {author} {\bibfnamefont {M.~A.}\ \bibnamefont {Forman}},
  \bibinfo {author} {\bibfnamefont {C.~W.}\ \bibnamefont {Smith}}, \bibinfo
  {author} {\bibfnamefont {B.~J.}\ \bibnamefont {Vasquez}}, \ and\ \bibinfo
  {author} {\bibfnamefont {J.~E.}\ \bibnamefont {Stawarz}},\ }\href {\doibase
  10.1098/rsta.2014.0150} {\bibfield  {journal} {\bibinfo  {journal}
  {Philosophical Transactions of the Royal Society A: Mathematical, Physical
  and Engineering Sciences}\ }\textbf {\bibinfo {volume} {373}},\ \bibinfo
  {pages} {20140150} (\bibinfo {year} {2015})}\BibitemShut {NoStop}%
\bibitem [{\citenamefont {{Banerjee}}\ \emph {et~al.}(2016)\citenamefont
  {{Banerjee}}, \citenamefont {{Hadid}}, \citenamefont {{Sahraoui}},\ and\
  \citenamefont {{Galtier}}}]{Banerjee2016}%
  \BibitemOpen
  \bibfield  {author} {\bibinfo {author} {\bibfnamefont {S.}~\bibnamefont
  {{Banerjee}}}, \bibinfo {author} {\bibfnamefont {L.~Z.}\ \bibnamefont
  {{Hadid}}}, \bibinfo {author} {\bibfnamefont {F.}~\bibnamefont {{Sahraoui}}},
  \ and\ \bibinfo {author} {\bibfnamefont {S.}~\bibnamefont {{Galtier}}},\
  }\href {\doibase 10.3847/2041-8205/829/2/L27} {\bibfield  {journal} {\bibinfo
   {journal} {Astrophys. J. Lett.}\ }\textbf {\bibinfo {volume} {829}},\
  \bibinfo {pages} {L27} (\bibinfo {year} {2016})}\BibitemShut {NoStop}%
\bibitem [{\citenamefont {Hadid}\ \emph {et~al.}(2018)\citenamefont {Hadid},
  \citenamefont {Sahraoui}, \citenamefont {Galtier},\ and\ \citenamefont
  {Huang}}]{Hadid2018}%
  \BibitemOpen
  \bibfield  {author} {\bibinfo {author} {\bibfnamefont {L.~Z.}\ \bibnamefont
  {Hadid}}, \bibinfo {author} {\bibfnamefont {F.}~\bibnamefont {Sahraoui}},
  \bibinfo {author} {\bibfnamefont {S.}~\bibnamefont {Galtier}}, \ and\
  \bibinfo {author} {\bibfnamefont {S.~Y.}\ \bibnamefont {Huang}},\ }\href
  {\doibase 10.1103/PhysRevLett.120.055102} {\bibfield  {journal} {\bibinfo
  {journal} {Phys. Rev. Lett.}\ }\textbf {\bibinfo {volume} {120}},\ \bibinfo
  {pages} {055102} (\bibinfo {year} {2018})}\BibitemShut {NoStop}%
\bibitem [{\citenamefont {Andr\'es}\ \emph {et~al.}(2019)\citenamefont
  {Andr\'es}, \citenamefont {Sahraoui}, \citenamefont {Galtier}, \citenamefont
  {Hadid}, \citenamefont {Ferrand},\ and\ \citenamefont {Huang}}]{Andres2019}%
  \BibitemOpen
  \bibfield  {author} {\bibinfo {author} {\bibfnamefont {N.}~\bibnamefont
  {Andr\'es}}, \bibinfo {author} {\bibfnamefont {F.}~\bibnamefont {Sahraoui}},
  \bibinfo {author} {\bibfnamefont {S.}~\bibnamefont {Galtier}}, \bibinfo
  {author} {\bibfnamefont {L.~Z.}\ \bibnamefont {Hadid}}, \bibinfo {author}
  {\bibfnamefont {R.}~\bibnamefont {Ferrand}}, \ and\ \bibinfo {author}
  {\bibfnamefont {S.~Y.}\ \bibnamefont {Huang}},\ }\href {\doibase
  10.1103/PhysRevLett.123.245101} {\bibfield  {journal} {\bibinfo  {journal}
  {Phys. Rev. Lett.}\ }\textbf {\bibinfo {volume} {123}},\ \bibinfo {pages}
  {245101} (\bibinfo {year} {2019})}\BibitemShut {NoStop}%
\bibitem [{\citenamefont {Hadid}\ \emph {et~al.}(2017)\citenamefont {Hadid},
  \citenamefont {Sahraoui},\ and\ \citenamefont {Galtier}}]{Hadid2017}%
  \BibitemOpen
  \bibfield  {author} {\bibinfo {author} {\bibfnamefont {L.~Z.}\ \bibnamefont
  {Hadid}}, \bibinfo {author} {\bibfnamefont {F.}~\bibnamefont {Sahraoui}}, \
  and\ \bibinfo {author} {\bibfnamefont {S.}~\bibnamefont {Galtier}},\ }\href
  {\doibase 10.3847/1538-4357/aa603f} {\bibfield  {journal} {\bibinfo
  {journal} {The Astrophysical Journal}\ }\textbf {\bibinfo {volume} {838}},\
  \bibinfo {pages} {9} (\bibinfo {year} {2017})}\BibitemShut {NoStop}%
\bibitem [{\citenamefont {Duchon}\ and\ \citenamefont {Robert}(1999)}]{Duchon}%
  \BibitemOpen
  \bibfield  {author} {\bibinfo {author} {\bibfnamefont {J.}~\bibnamefont
  {Duchon}}\ and\ \bibinfo {author} {\bibfnamefont {R.}~\bibnamefont
  {Robert}},\ }\href {\doibase 10.1088/0951-7715/13/1/312} {\bibfield
  {journal} {\bibinfo  {journal} {Nonlinearity}\ }\textbf {\bibinfo {volume}
  {13}},\ \bibinfo {pages} {249} (\bibinfo {year} {1999})}\BibitemShut
  {NoStop}%
\bibitem [{\citenamefont {Leray}(1934)}]{Leray34}%
  \BibitemOpen
  \bibfield  {author} {\bibinfo {author} {\bibfnamefont {J.}~\bibnamefont
  {Leray}},\ }\href {\doibase 10.1007/BF02547354} {\bibfield  {journal}
  {\bibinfo  {journal} {Acta Math.}\ }\textbf {\bibinfo {volume} {63}},\
  \bibinfo {pages} {193} (\bibinfo {year} {1934})}\BibitemShut {NoStop}%
\bibitem [{\citenamefont {Burlaga}\ and\ \citenamefont
  {Goldstein}(1984)}]{Burlaga1984}%
  \BibitemOpen
  \bibfield  {author} {\bibinfo {author} {\bibfnamefont {L.~F.}\ \bibnamefont
  {Burlaga}}\ and\ \bibinfo {author} {\bibfnamefont {M.~L.}\ \bibnamefont
  {Goldstein}},\ }\href {\doibase https://doi.org/10.1029/JA089iA08p06813}
  {\bibfield  {journal} {\bibinfo  {journal} {Journal of Geophysical Research:
  Space Physics}\ }\textbf {\bibinfo {volume} {89}},\ \bibinfo {pages} {6813}
  (\bibinfo {year} {1984})}\BibitemShut {NoStop}%
\bibitem [{\citenamefont {{Roberts}}\ and\ \citenamefont
  {{Goldstein}}(1987)}]{Roberts1987}%
  \BibitemOpen
  \bibfield  {author} {\bibinfo {author} {\bibfnamefont {D.~A.}\ \bibnamefont
  {{Roberts}}}\ and\ \bibinfo {author} {\bibfnamefont {M.~L.}\ \bibnamefont
  {{Goldstein}}},\ }\href {\doibase 10.1029/JA092iA09p10105} {\bibfield
  {journal} {\bibinfo  {journal} {J. Geophys. Res}\ }\textbf {\bibinfo {volume}
  {92}},\ \bibinfo {pages} {10105} (\bibinfo {year} {1987})}\BibitemShut
  {NoStop}%
\bibitem [{\citenamefont {Dubrulle}(2019)}]{Dubrulle19}%
  \BibitemOpen
  \bibfield  {author} {\bibinfo {author} {\bibfnamefont {B.}~\bibnamefont
  {Dubrulle}},\ }\href {\doibase 10.1017/jfm.2019.98} {\bibfield  {journal}
  {\bibinfo  {journal} {Journal of Fluid Mechanics}\ }\textbf {\bibinfo
  {volume} {867}},\ \bibinfo {pages} {P1} (\bibinfo {year} {2019})}\BibitemShut
  {NoStop}%
\bibitem [{\citenamefont {Galtier}(2018)}]{G18}%
  \BibitemOpen
  \bibfield  {author} {\bibinfo {author} {\bibfnamefont {S.}~\bibnamefont
  {Galtier}},\ }\href {\doibase 10.1088/1751-8121/aabbb5} {\bibfield  {journal}
  {\bibinfo  {journal} {Journal of Physics A: Mathematical and Theoretical}\
  }\textbf {\bibinfo {volume} {51}},\ \bibinfo {pages} {205501} (\bibinfo
  {year} {2018})}\BibitemShut {NoStop}%
\bibitem [{\citenamefont {Bec}\ and\ \citenamefont {Khanin}(2007)}]{Bec2007}%
  \BibitemOpen
  \bibfield  {author} {\bibinfo {author} {\bibfnamefont {J.}~\bibnamefont
  {Bec}}\ and\ \bibinfo {author} {\bibfnamefont {K.}~\bibnamefont {Khanin}},\
  }\href {\doibase https://doi.org/10.1016/j.physrep.2007.04.002} {\bibfield
  {journal} {\bibinfo  {journal} {Physics Reports}\ }\textbf {\bibinfo {volume}
  {447}},\ \bibinfo {pages} {1 } (\bibinfo {year} {2007})}\BibitemShut
  {NoStop}%
\bibitem [{\citenamefont {Yanase}(1997)}]{Yanase97}%
  \BibitemOpen
  \bibfield  {author} {\bibinfo {author} {\bibfnamefont {S.}~\bibnamefont
  {Yanase}},\ }\href {\doibase 10.1063/1.872190} {\bibfield  {journal}
  {\bibinfo  {journal} {Physics of Plasmas}\ }\textbf {\bibinfo {volume} {4}},\
  \bibinfo {pages} {1010} (\bibinfo {year} {1997})}\BibitemShut {NoStop}%
\bibitem [{\citenamefont {Thomas}(1968)}]{Thomas1968}%
  \BibitemOpen
  \bibfield  {author} {\bibinfo {author} {\bibfnamefont {J.~H.}\ \bibnamefont
  {Thomas}},\ }\href {\doibase 10.1063/1.1692092} {\bibfield  {journal}
  {\bibinfo  {journal} {The Physics of Fluids}\ }\textbf {\bibinfo {volume}
  {11}},\ \bibinfo {pages} {1245} (\bibinfo {year} {1968})}\BibitemShut
  {NoStop}%
\bibitem [{\citenamefont {{Galtier}}\ and\ \citenamefont
  {{Pouquet}}(1998)}]{Galtier1998}%
  \BibitemOpen
  \bibfield  {author} {\bibinfo {author} {\bibfnamefont {S.}~\bibnamefont
  {{Galtier}}}\ and\ \bibinfo {author} {\bibfnamefont {A.}~\bibnamefont
  {{Pouquet}}},\ }\href {\doibase 10.1023/A:1005056102064} {\bibfield
  {journal} {\bibinfo  {journal} {Solar Phys.}\ }\textbf {\bibinfo {volume}
  {179}},\ \bibinfo {pages} {141} (\bibinfo {year} {1998})}\BibitemShut
  {NoStop}%
\bibitem [{\citenamefont {Suzuki}\ and\ \citenamefont
  {Inutsuka}(2005)}]{Suzuki2005}%
  \BibitemOpen
  \bibfield  {author} {\bibinfo {author} {\bibfnamefont {T.~K.}\ \bibnamefont
  {Suzuki}}\ and\ \bibinfo {author} {\bibfnamefont {S.-I.}\ \bibnamefont
  {Inutsuka}},\ }\href {\doibase 10.1086/497536} {\bibfield  {journal}
  {\bibinfo  {journal} {The Astrophysical Journal}\ }\textbf {\bibinfo {volume}
  {632}},\ \bibinfo {pages} {L49} (\bibinfo {year} {2005})}\BibitemShut
  {NoStop}%
\bibitem [{\citenamefont {Eyink}(2019)}]{EyinkNotes}%
  \BibitemOpen
  \bibfield  {author} {\bibinfo {author} {\bibfnamefont {G.}~\bibnamefont
  {Eyink}},\ }\href {http://www.ams.jhu.edu/~eyink/Turbulence/notes.html}
  {\enquote {\bibinfo {title} {Small-scale intermittency and anomalous
  scaling},}\ } (\bibinfo {year} {2019})\BibitemShut {NoStop}%
\bibitem [{\citenamefont {Soluyan}\ and\ \citenamefont
  {Khokhlov}(1961)}]{Khokhlov61}%
  \BibitemOpen
  \bibfield  {author} {\bibinfo {author} {\bibfnamefont {S.}~\bibnamefont
  {Soluyan}}\ and\ \bibinfo {author} {\bibfnamefont {R.}~\bibnamefont
  {Khokhlov}},\ }\href@noop {} {\bibfield  {journal} {\bibinfo  {journal}
  {Vestnik Moscow State Univ., Phys. Astron}\ }\textbf {\bibinfo {volume}
  {3}},\ \bibinfo {pages} {52} (\bibinfo {year} {1961})}\BibitemShut {NoStop}%
\bibitem [{\citenamefont {{Pine}}\ \emph
  {et~al.}(2020{\natexlab{a}})\citenamefont {{Pine}}, \citenamefont {{Smith}},
  \citenamefont {{Hollick}}, \citenamefont {{Argall}}, \citenamefont
  {{Vasquez}}, \citenamefont {{Isenberg}}, \citenamefont {{Schwadron}},
  \citenamefont {{Joyce}}, \citenamefont {{Sok{\'o}{\l}}}, \citenamefont
  {{Bzowski}}, \citenamefont {{Kubiak}},\ and\ \citenamefont
  {{McLaurin}}}]{Pine2020}%
  \BibitemOpen
  \bibfield  {author} {\bibinfo {author} {\bibfnamefont {Z.~B.}\ \bibnamefont
  {{Pine}}}, \bibinfo {author} {\bibfnamefont {C.~W.}\ \bibnamefont {{Smith}}},
  \bibinfo {author} {\bibfnamefont {S.~J.}\ \bibnamefont {{Hollick}}}, \bibinfo
  {author} {\bibfnamefont {M.~R.}\ \bibnamefont {{Argall}}}, \bibinfo {author}
  {\bibfnamefont {B.~J.}\ \bibnamefont {{Vasquez}}}, \bibinfo {author}
  {\bibfnamefont {P.~A.}\ \bibnamefont {{Isenberg}}}, \bibinfo {author}
  {\bibfnamefont {N.~A.}\ \bibnamefont {{Schwadron}}}, \bibinfo {author}
  {\bibfnamefont {C.~J.}\ \bibnamefont {{Joyce}}}, \bibinfo {author}
  {\bibfnamefont {J.~M.}\ \bibnamefont {{Sok{\'o}{\l}}}}, \bibinfo {author}
  {\bibfnamefont {M.}~\bibnamefont {{Bzowski}}}, \bibinfo {author}
  {\bibfnamefont {M.~A.}\ \bibnamefont {{Kubiak}}}, \ and\ \bibinfo {author}
  {\bibfnamefont {M.~L.}\ \bibnamefont {{McLaurin}}},\ }\href {\doibase
  10.3847/1538-4357/abab12} {\bibfield  {journal} {\bibinfo  {journal}
  {Astrophys. J.}\ }\textbf {\bibinfo {volume} {900}},\ \bibinfo {pages} {94}
  (\bibinfo {year} {2020}{\natexlab{a}})}\BibitemShut {NoStop}%
\bibitem [{\citenamefont {{Pine}}\ \emph
  {et~al.}(2020{\natexlab{b}})\citenamefont {{Pine}}, \citenamefont {{Smith}},
  \citenamefont {{Hollick}}, \citenamefont {{Argall}}, \citenamefont
  {{Vasquez}}, \citenamefont {{Isenberg}}, \citenamefont {{Schwadron}},
  \citenamefont {{Joyce}}, \citenamefont {{Sok{\'o}{\l}}}, \citenamefont
  {{Bzowski}}, \citenamefont {{Kubiak}}, \citenamefont {{Hamilton}},
  \citenamefont {{McLaurin}},\ and\ \citenamefont {{Leamon}}}]{Pine2020b}%
  \BibitemOpen
  \bibfield  {author} {\bibinfo {author} {\bibfnamefont {Z.~B.}\ \bibnamefont
  {{Pine}}}, \bibinfo {author} {\bibfnamefont {C.~W.}\ \bibnamefont {{Smith}}},
  \bibinfo {author} {\bibfnamefont {S.~J.}\ \bibnamefont {{Hollick}}}, \bibinfo
  {author} {\bibfnamefont {M.~R.}\ \bibnamefont {{Argall}}}, \bibinfo {author}
  {\bibfnamefont {B.~J.}\ \bibnamefont {{Vasquez}}}, \bibinfo {author}
  {\bibfnamefont {P.~A.}\ \bibnamefont {{Isenberg}}}, \bibinfo {author}
  {\bibfnamefont {N.~A.}\ \bibnamefont {{Schwadron}}}, \bibinfo {author}
  {\bibfnamefont {C.~J.}\ \bibnamefont {{Joyce}}}, \bibinfo {author}
  {\bibfnamefont {J.~M.}\ \bibnamefont {{Sok{\'o}{\l}}}}, \bibinfo {author}
  {\bibfnamefont {M.}~\bibnamefont {{Bzowski}}}, \bibinfo {author}
  {\bibfnamefont {M.~A.}\ \bibnamefont {{Kubiak}}}, \bibinfo {author}
  {\bibfnamefont {K.~E.}\ \bibnamefont {{Hamilton}}}, \bibinfo {author}
  {\bibfnamefont {M.~L.}\ \bibnamefont {{McLaurin}}}, \ and\ \bibinfo {author}
  {\bibfnamefont {R.~J.}\ \bibnamefont {{Leamon}}},\ }\href {\doibase
  10.3847/1538-4357/abab0f} {\bibfield  {journal} {\bibinfo  {journal}
  {Astrophys. J.}\ }\textbf {\bibinfo {volume} {900}},\ \bibinfo {pages} {92}
  (\bibinfo {year} {2020}{\natexlab{b}})}\BibitemShut {NoStop}%
\bibitem [{\citenamefont {{Malara}}\ \emph {et~al.}(1999)\citenamefont
  {{Malara}}, \citenamefont {{Primavera}},\ and\ \citenamefont
  {{Veltri}}}]{Malara1999}%
  \BibitemOpen
  \bibfield  {author} {\bibinfo {author} {\bibfnamefont {F.}~\bibnamefont
  {{Malara}}}, \bibinfo {author} {\bibfnamefont {L.}~\bibnamefont
  {{Primavera}}}, \ and\ \bibinfo {author} {\bibfnamefont {P.}~\bibnamefont
  {{Veltri}}},\ }\href {\doibase 10.1103/PhysRevE.59.6023} {\bibfield
  {journal} {\bibinfo  {journal} {Phys. Rev. E}\ }\textbf {\bibinfo {volume}
  {59}},\ \bibinfo {pages} {6023} (\bibinfo {year} {1999})}\BibitemShut
  {NoStop}%
\bibitem [{\citenamefont {{Matthaeus}}\ and\ \citenamefont
  {{Goldstein}}(1982)}]{Matthaeus1982}%
  \BibitemOpen
  \bibfield  {author} {\bibinfo {author} {\bibfnamefont {W.~H.}\ \bibnamefont
  {{Matthaeus}}}\ and\ \bibinfo {author} {\bibfnamefont {M.~L.}\ \bibnamefont
  {{Goldstein}}},\ }\href {\doibase 10.1029/JA087iA08p06011} {\bibfield
  {journal} {\bibinfo  {journal} {J. Geophys. Res.}\ }\textbf {\bibinfo
  {volume} {87}},\ \bibinfo {pages} {6011} (\bibinfo {year}
  {1982})}\BibitemShut {NoStop}%
\bibitem [{\citenamefont {{Burlaga}}\ and\ \citenamefont
  {{Mish}}(1987)}]{Burlaga1987}%
  \BibitemOpen
  \bibfield  {author} {\bibinfo {author} {\bibfnamefont {L.~F.}\ \bibnamefont
  {{Burlaga}}}\ and\ \bibinfo {author} {\bibfnamefont {W.~H.}\ \bibnamefont
  {{Mish}}},\ }\href {\doibase 10.1029/JA092iA02p01261} {\bibfield  {journal}
  {\bibinfo  {journal} {J. Geophys. Res.}\ }\textbf {\bibinfo {volume} {92}},\
  \bibinfo {pages} {1261} (\bibinfo {year} {1987})}\BibitemShut {NoStop}%
\bibitem [{\citenamefont {{Burlaga}}\ \emph {et~al.}(1987)\citenamefont
  {{Burlaga}}, \citenamefont {{Ness}},\ and\ \citenamefont
  {{McDonald}}}]{Burlaga1987b}%
  \BibitemOpen
  \bibfield  {author} {\bibinfo {author} {\bibfnamefont {L.~F.}\ \bibnamefont
  {{Burlaga}}}, \bibinfo {author} {\bibfnamefont {N.~F.}\ \bibnamefont
  {{Ness}}}, \ and\ \bibinfo {author} {\bibfnamefont {F.~B.}\ \bibnamefont
  {{McDonald}}},\ }\href {\doibase 10.1029/JA092iA12p13647} {\bibfield
  {journal} {\bibinfo  {journal} {J. Geophys. Res.}\ }\textbf {\bibinfo
  {volume} {92}},\ \bibinfo {pages} {13647} (\bibinfo {year}
  {1987})}\BibitemShut {NoStop}%
\bibitem [{\citenamefont {{Miceli}}\ \emph {et~al.}(2019)\citenamefont
  {{Miceli}}, \citenamefont {{Orlando}}, \citenamefont {{Burrows}},
  \citenamefont {{Frank}}, \citenamefont {{Argiroffi}}, \citenamefont
  {{Reale}}, \citenamefont {{Peres}}, \citenamefont {{Petruk}},\ and\
  \citenamefont {{Bocchino}}}]{Miceli2019}%
  \BibitemOpen
  \bibfield  {author} {\bibinfo {author} {\bibfnamefont {M.}~\bibnamefont
  {{Miceli}}}, \bibinfo {author} {\bibfnamefont {S.}~\bibnamefont {{Orlando}}},
  \bibinfo {author} {\bibfnamefont {D.~N.}\ \bibnamefont {{Burrows}}}, \bibinfo
  {author} {\bibfnamefont {K.~A.}\ \bibnamefont {{Frank}}}, \bibinfo {author}
  {\bibfnamefont {C.}~\bibnamefont {{Argiroffi}}}, \bibinfo {author}
  {\bibfnamefont {F.}~\bibnamefont {{Reale}}}, \bibinfo {author} {\bibfnamefont
  {G.}~\bibnamefont {{Peres}}}, \bibinfo {author} {\bibfnamefont
  {O.}~\bibnamefont {{Petruk}}}, \ and\ \bibinfo {author} {\bibfnamefont
  {F.}~\bibnamefont {{Bocchino}}},\ }\href {\doibase 10.1038/s41550-018-0677-8}
  {\bibfield  {journal} {\bibinfo  {journal} {Nature Astronomy}\ }\textbf
  {\bibinfo {volume} {3}},\ \bibinfo {pages} {236} (\bibinfo {year}
  {2019})}\BibitemShut {NoStop}%
\end{thebibliography}%
\end{document}